\documentclass[iop,apj,appendixfloats]{emulateapj}
\usepackage{apjfonts}

\gdef\kms{km\,s$^{-1}$}
\gdef\msun{$M_{\odot}$}
\lefthead{van Dokkum et al.}
\righthead{}
\slugcomment{Astrophysical Journal, in press}
\begin{document}

\title{Dense cores in galaxies out to $z=2.5$ in SDSS, UltraVISTA,
and the Five 3D-HST/CANDELS Fields}

\author{Pieter G.\ van Dokkum\altaffilmark{1},
Rachel Bezanson\altaffilmark{2},
Arjen van der Wel\altaffilmark{3},
Erica June Nelson\altaffilmark{1},
Ivelina Momcheva\altaffilmark{1},
Rosalind E.\ Skelton\altaffilmark{4},
Katherine E.\ Whitaker\altaffilmark{5},
Gabriel Brammer\altaffilmark{6},
Charlie Conroy\altaffilmark{7},
Natascha M.\ F\"orster Schreiber\altaffilmark{8},
Mattia Fumagalli\altaffilmark{9},
Mariska Kriek\altaffilmark{10},
Ivo Labb\'e\altaffilmark{4},
Joel Leja\altaffilmark{1},
Danilo Marchesini\altaffilmark{11},
Adam Muzzin\altaffilmark{4},
Pascal Oesch\altaffilmark{1},
Stijn Wuyts\altaffilmark{9}
}

\altaffiltext{1}
{Department of Astronomy, Yale University, New Haven, CT 06511, USA}
\altaffiltext{2}
{Steward Observatory, University of Arizona, 933 North Cherry Avenue,
Tucson, AZ 85721, USA}
\altaffiltext{3}
{Max Planck Institute for Astronomy (MPIA), K\"onigstuhl 17, D-69117,
Heidelberg, Germany}
\altaffiltext{4}
{Department of Astronomy, University of Cape Town, Private Bag X3, Rondebosch 7701, South Africa}
\altaffiltext{5}
{Astrophysics Science Division, Goddard Space Center, Greenbelt,
MD 20771, USA}
\altaffiltext{6}
{Space Telescope Science Institute, Baltimore, MD 21218, USA}
\altaffiltext{7}
{Department of Astronomy \& Astrophysics, University of California, Santa Cruz, CA, USA}
\altaffiltext{8}
{Max-Planck-Institut f\"ur extraterrestrische Physik, Giessenbachstrasse,
D-85748 Garching, Germany}
\altaffiltext{9}
{Leiden Observatory, Leiden University, Leiden, The Netherlands}
\altaffiltext{10}
{Department of Astronomy, University of California, Berkeley, CA 94720, USA}
\altaffiltext{11}
{Department of Physics and Astronomy, Tufts University, Medford, MA 02155,
USA}

\begin{abstract}

The dense interiors of massive galaxies are among the most
intriguing environments in the Universe. In this paper we ask
when these dense cores were formed and determine how
galaxies gradually assembled around them. We select
galaxies that have a stellar mass $>3\times 10^{10}$\,\msun\
inside $r=1$\,kpc  out to $z=2.5$, using
the 3D-HST survey and data at low redshift.
Remarkably, the  number
density of galaxies with dense cores appears to have decreased from
$z=2.5$ to the present. This decrease is probably mostly
due to stellar mass loss and
the resulting adiabatic expansion, with some contribution from merging.
We infer
that dense cores were mostly formed at $z>2.5$, consistent with
their largely quiescent stellar populations. While
the cores appear to form early, the galaxies
in which they reside show strong evolution:
their total
masses increase by
a factor of $2-3$ from $z=2.5$ to $z=0$ and their effective
radii increase by a factor of $5-6$.
As a result, the contribution of dense cores
to the total mass of the galaxies in which they reside
decreases from $\sim 50$\,\%  at $z=2.5$ to
$\sim 15$\,\% at $z=0$.
Because
of their early formation, the contribution
of dense cores to the total stellar mass budget of the Universe is a strong
function of redshift. The stars in
cores with $M_{\rm 1\,kpc}>3\times 10^{10}$\,\msun\
make up $\sim 0.1$\,\% of the stellar mass density of the Universe
today but 10\,\% -- 20\,\% at $z\sim 2$, depending on their IMF.
The formation of these cores
required the conversion of $\sim 10^{11}$\,\msun\ of gas into
stars within $\sim 1\,$kpc, while preventing significant
star formation
at larger radii.

\end{abstract}


\section{Introduction}

The central regions of massive elliptical
galaxies such as NGC\,1399 and NGC\,4472
are  different from any environment seen in galaxies
such as the Milky Way.
The mean stellar densities 
are $\sim 10$\,\msun\,pc$^{-3}$ in the central
kpc, and their velocity dispersions reach or even exceed
$\sim 300$\,\kms.  The stellar populations are old,
metal rich, and strongly $\alpha$-enhanced, indicating that the stars
were formed early in a short, intense period of star formation
({Franx} \& {Illingworth} 1990; {Worthey}, {Faber}, \&  {Gonzalez} 1992; {Davies}, {Sadler}, \& {Peletier} 1993; {Kuntschner} {et~al.} 2001, 2010, and many other studies).
Star formation in these central regions likely took place under
very different
physical conditions than
those in the present-day disk of the Milky Way, possibly leading
to a bottom-heavy IMF with an excess of low mass stars compared
to the Milky Way IMF ({van Dokkum} \& {Conroy} 2010; {Treu} {et~al.} 2010; {Krumholz} 2011; {Cappellari} {et~al.} 2012; {Conroy} \& {van Dokkum} 2012; {Hopkins} 2013).
These dense centers also host the most massive black
holes in the Universe ({Magorrian} {et~al.} 1998; {Ferrarese} \& {Merritt} 2000; {Gebhardt} {et~al.} 2000),
which probably accreted  most of their mass
during the peak star formation epoch.
Despite their high star formation efficiency in the past,
dense regions are hostile to star formation today: quiescence
correlates well with velocity dispersion and with stellar surface
density ({Kauffmann} {et~al.} 2003; {Franx} {et~al.} 2008; {Wake}, {van Dokkum}, \& {Franx} 2012; {Bell} {et~al.} 2012).

The dense interiors of massive
galaxies account for only a small
fraction of the total stellar mass in the present-day
Universe, but given the old
ages of their stars this fraction is expected to increase with
redshift. In fact, the formation of the dense central
parts of elliptical
galaxies may preceed the assembly of the rest of the galaxies. Many
quiescent galaxies at $z=1.5 - 2.5$
are much more compact than nearby galaxies of the same mass
(e.g., {Daddi} {et~al.} 2005; {Trujillo} {et~al.} 2006; {van Dokkum} {et~al.} 2008; {Cimatti} {et~al.} 2008; {Damjanov} {et~al.} 2009; {Williams} {et~al.} 2010), and
as first shown by {Bezanson} {et~al.} (2009) the central
densities of the compact high redshift galaxies are broadly
similar to those
of massive elliptical galaxies today. 
This is consistent with the idea that
massive galaxies have grown inside-out since $z\sim 2$,
with their cores forming at higher redshift
and their outer envelopes building up slowly through star formation,
minor mergers, or other
processes (e.g., {Loeb} \& {Peebles} 2003; {Bezanson} {et~al.} 2009; {Naab}, {Johansson}, \& {Ostriker} 2009; {van Dokkum} {et~al.} 2010; {Hopkins} {et~al.} 2010; {Oser} {et~al.} 2010; Feldmann et al.\ 2010; {Szomoru} {et~al.} 2013).


In this paper we focus exclusively on these dense central
regions of massive galaxies: we ask what
their number density is, what their contribution is to the overall stellar
mass density, and how the galaxies that they are part of were built up
around them. In practice, we select
galaxies out to $z=2.5$
that have $\log M_{\rm 1\,kpc}\gtrsim 10.5$, that is,
a stellar mass exceeding
$3.2 \times 10^{10}$\,\msun\
within a sphere of radius $r=1$\,kpc.\footnote{We refer to the region within
this radius as the ``core'' throughout this paper, realizing that
this may cause confusion. The same term has been used extensively
in the literature to describe the surface density profile of
early-type galaxies on much smaller scales (e.g., {Faber} {et~al.} 1997).}
We do not limit the sample
to quiescent galaxies but select all objects that satisfy this
stellar density criterion.
Our approach is different from studies of the properties of
galaxies at fixed total stellar mass, or fixed 
number density. In fact, as 
we show in \S\,\ref{ndens.sec} the evolution of the
number density of galaxies with
$\log M_{\rm 1\,kpc}>10.5$ is different from
that of the general population of massive galaxies.
Our study is more closely
related to the work of {Bezanson} {et~al.} (2011) on the
evolution of the velocity dispersion function; Bezanson et al.\
converted observed effective radii, stellar masses, and {Sersic} (1968)
indices to velocity dispersions whereas we convert the same parameters
to a stellar mass within a physical radius of 1\,kpc.

In this paper we do not make any a priori selection on star
formation rate or galaxy size. Nevertheless, this paper has implications
for the evolution of
massive quiescent galaxies at $z\sim 2$. It is
generally thought that these galaxies have grown substantially in
size over the past 10 Gyr, but this interpretation is complicated by
the fact that the number density of quiescent galaxies has increased
by an order of magnitude over this time period
({Brammer} {et~al.} 2011; {Cassata} {et~al.} 2013; {Muzzin} {et~al.} 2013a). As discussed by
{van Dokkum} {et~al.} (2008),
{van der Wel} {et~al.} (2009), Trujillo et al.\ (2011),
{Newman} {et~al.} (2012), {Carollo} {et~al.} (2013),
{Szomoru} {et~al.} (2013), and others,
the evolution of the mass-size relation of quiescent galaxies
could be partially driven by the continuous
addition of large, recently quenched star-forming galaxies, in which case 
the growth of individual quiescent galaxies would be 
smaller than that of the population. Some studies have even
suggested that compact quiescent galaxies barely evolve at all
(e.g., {Poggianti} {et~al.} 2013). As we show in \S\,\ref{ndens.sec} the
evolution of galaxies with dense cores appears to require substantial
evolution in the sizes and masses of individual compact galaxies
after $z\sim 2$.

The paper is structured as follows. In \S\,\ref{data.sec} we describe
the sources of data that are used.  In \S\,\ref{select.sec} the
selection of galaxies with dense cores is described. Sections
\ref{ndens.sec} and \ref{buildup.sec} form the heart of the
paper. In \S\,\ref{ndens.sec} the
``core mass function'' is discussed, that is, the
number of galaxies as a function of their mass within 1 kpc. This
section also presents the evolution of the cumulative number density
of galaxies with $\log M_{\rm 1\,kpc}>10.5$, and interprets the
evolution in the context of various physical processes. 
Finally, it
places the total stellar mass locked up
in dense cores in the context
of the evolving stellar mass density of the Universe.
In \S\,\ref{buildup.sec} the properties of galaxies
that have dense cores are analyzed;
here we show that the core-hosting galaxies likely
evolved significantly since $z\sim 2$, increasing
both their total mass and (particularly) their effective radii.
We also discuss the nature of star forming galaxies with dense
cores.
The paper is summarized in \S\,\ref{conclusions.sec}.

\section{Data}
\label{data.sec}

\subsection{The 3D-HST Survey and Catalog}
We use the imaging, spectroscopy, and catalogs from the 3D-HST survey
({Brammer} {et~al.} 2012). 3D-HST is an HST Treasury program that has provided
WFC3/G141 grism spectroscopy over four of the five extra-galactic fields imaged
by the CANDELS survey ({Grogin} {et~al.} 2011; {Koekemoer} {et~al.} 2011).
Including archival data on GOODS-North from program GO-11600 (PI: Weiner)
approximately 80\,\% of the CANDELS area is covered by grism
observations.

In addition to analyzing the grism spectroscopy, the 3D-HST project reduced all
the CANDELS WFC3 imaging, and has constructed photometric catalogs in
the five CANDELS fields using
publicly available ground- and space-based photometry from $0.3\,\mu$m --
$8.0\,\mu$m. This multi-wavelength photometry aids in the interpretation
of the grism spectroscopy and
is obviously valuable
in its own right.
The complete
CANDELS + 3D-HST datasets in all five CANDELS fields are included
in v4.1 of the catalogs.\footnote{http://3dhst.research.yale.edu/Home.html}
Here we give
a brief description of the 3D-HST data products; the full catalogs
are presented and described in {Skelton} {et~al.} (2014).

Redshifts were measured from a combination of the
$U$--IRAC photometric data and the WFC3/G141 grism spectra,
using a modified version of the EAZY code ({Brammer}, {van Dokkum}, \& {Coppi} 2008) as described
in {Brammer} {et~al.} (2013).
Comparisons to ground-based spectroscopic redshifts
suggest an accuracy of $0.003-0.005$ in $\Delta z/(1+z)$ for galaxies
with $H\lesssim 23$; this seems to be borne out by stacking analyses of
galaxies without a previously measured
redshift (see {Whitaker} {et~al.} 2013).
For faint galaxies, and in areas of the CANDELS fields that do not
have grism coverage, we use the photometric redshift instead.
Structural parameters in $J_{125}$ and $H_{160}$
were measured using GALFIT ({Peng} {et~al.} 2002), as
described in {van der Wel} {et~al.} (2014). The {Sersic} (1968)
parameters measured from the 3D-HST mosaics
are in excellent agreement with those measured from the CANDELS mosaics
({van der Wel} {et~al.} 2012) for the same objects. Stellar masses were measured
from the photometric data using the FAST code ({Kriek} {et~al.} 2009b), assuming
a {Chabrier} (2003) IMF. 

Excluding the areas surrounding bright stars and regions
with little WFC3 exposure time (such as the edge of
the field), the five fields cover a total of 896\,arcmin$^2$
in version 4.1 of our catalogs:
192.4\,arcmin$^2$ in AEGIS, 183.9\,arcmin$^2$ in COSMOS, 157.8\,arcmin$^2$
in GOODS-North, 171.0\,arcmin$^2$ in GOODS-South, and 191.2\,arcmin$^2$
in UDS (see {Grogin} {et~al.} 2011; {Skelton} {et~al.} 2014, for a detailed description of these fields).

\subsection{The UltraVISTA and ``Z\"urich'' Catalogs in the COSMOS Field}

Dense cores are rare and large volumes are required to measure their
number density accurately. As we show in \S\,\ref{ndens.sec} the number
density of galaxies with $\log(M_{\rm 1\,kpc})>10.5$ is less than
$10^{-4}$\,Mpc$^{-3}$ at low redshift.
The volume probed by the 3D-HST/CANDELS survey is  only $1.7\times
10^5$\,Mpc$^3$ at $0<z<0.5$, which means of order 10 galaxies
can be expected. As this redshift range covers approximately
half of the time elapsed since $z=2.5$, we augment the 3D-HST/CANDELS
survey with the wide-field UltraVISTA survey in the COSMOS field.

We use the deep $K$-selected catalog of {Muzzin} {et~al.} (2013b),
which is based on the data described in McCracken et al.\ (2012). The
datasets and procedures used by Muzzin et al.\ are similar
to those used in our analysis of the CANDELS/3D-HST data. In
particular, photometric redshifts and stellar masses were derived using
the same software and assumptions, which means they can be combined
with our higher redshift data.
This catalog was matched to the ``Z\"urich Structure and Morphology''
catalog v1.0 ({Sargent} {et~al.} 2007), which contains structural
parameters of objects 
in the COSMOS field to a limiting magnitude of $I=22.5$.
Sizes and Sersic indices
were derived from the
HST/ACS $I_{814}$ imaging in this field
({Scoville} {et~al.} 2007), using the GIM2D
software ({Simard} {et~al.} 2002).
The total area that is covered by both UltraVISTA and ACS
is 1.54\,degree$^2$, a factor
of 6 larger than the 3D-HST/CANDELS survey.
We note that the COSMOS objects studied in this paper, galaxies with
dense cores at $0<z<0.5$, are bright and far removed from the
limits of the data. Also, the $I_{814}$
images are well matched in rest-frame wavelength to the $J_{125}$ and
$H_{160}$ that are used at higher redshifts.

\subsection{The Sloan Digital Sky Survey}

We also make use of a local sample, drawn from the 7th data
release of the Sloan Digital Sky Survey
(SDSS). This sample is described in {Bezanson} {et~al.} (2013); it is based
on structural
parameters measured by {Simard} {et~al.} (2011) and $M/L$ ratios from the MPA-JHU
catalog ({Brinchmann} {et~al.} 2004). The following
cuts were applied: keep\_flag=1, z\_warning=0, sciencePrimary=1,
and a requirement that structural parameters
and masses are available from {Simard} {et~al.} (2011)
and the MPA-JHU catalog respectively.\footnote{{Bezanson} {et~al.} (2013) also
required the error in the measured stellar velocity dispersion to be
below 10\,\%. This
additional constraint was not applied here.}
An effective area of 8032\,degree$^2$ was assumed (see {Simard} {et~al.} 2011).
Unfortunately,
we have no direct test to assess whether the SDSS masses and sizes
are on the same system as the 3D-HST data. There is no evidence to
the contrary: as shown in {Bezanson} {et~al.} (2013) there
are no obvious systematic differences in sizes or masses between
this SDSS sample and the extrapolation to $z=0$ 
of distant galaxy samples analyzed in the same way as done here.

\begin{figure*}[hbtp]
\epsfxsize=16cm
\begin{center}
\epsffile[66 310 476 594]{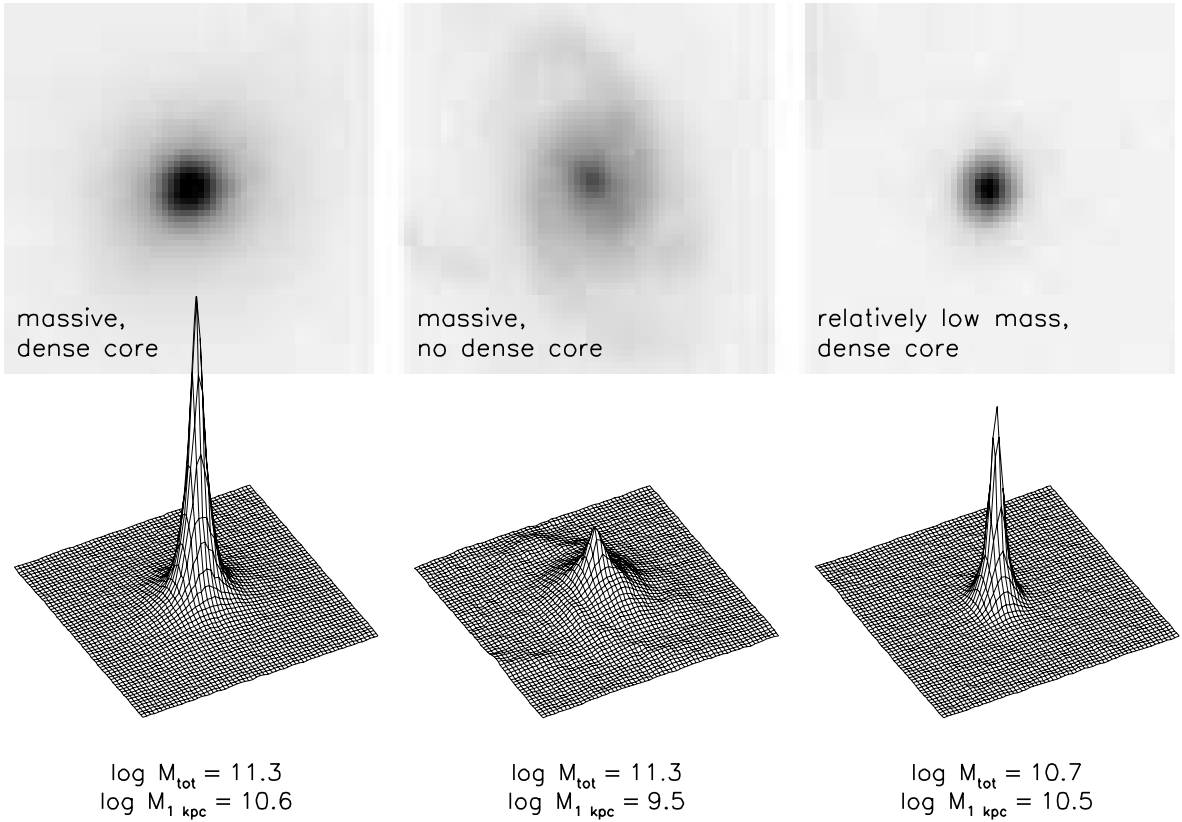}
\end{center}
\caption{\small
Illustration of the distinction between central mass and total mass.
The panels show HST/WFC3 $H_{160}$
images of three galaxies at $z\sim 1$; each panel spans
30\,kpc\,$\times$\,30\,kpc. The galaxy on the left has a high total
mass  and a high core mass; the middle galaxy has a high total mass
but a low core mass; and the galaxy on the right is
compact with a relatively low total mass and a high core mass.
\label{examples.fig}}
\end{figure*}

\section{Core Masses}
\label{select.sec}

\subsection{Selection at $0.5<z<2.5$}

The parent sample constitutes all galaxies in the five 3D-HST/CANDELS fields
that have a photometric ``use'' flag of 1
(see Skelton et al.\ 2014)
and a structural parameter fit flag of 0 or 1 (see {van der Wel} {et~al.} 2012,
2014).
For all these galaxies (essentially the entire 3D-HST v4.0 catalog)
we calculated the stellar mass that is contained within a
radius of 1\,kpc. Following {Bezanson} {et~al.} (2009) we first deprojected the
best-fitting Sersic profile using an Abel Transform:
\begin{equation}
\rho(x) = \frac{b_n}{\pi}\frac{I_0}{r_e} x^{1/n-1} \int_1^{\infty}
\frac{\exp\left(-b_n x^{1/n} t\right)}{\sqrt{t^{2n}-1}}
dt,
\label{abel.eq}
\end{equation}
with $\rho$ the 3D luminosity density in a particular filter,
$x\equiv r/r_e$, $r_e$ the circularized effective radius,
$n$ the Sersic index, and $b_n$ the $n$-dependent normalization parameter
of the Sersic profile (see, e.g., {Peng} {et~al.} 2002). 
This deprojection is important as the projected mass within 1\,kpc
is influenced by the properties of the galaxy at larger radius. In particular,
for a given density (in M$_{\odot}$\,kpc$^{-3}$)
within a sphere of radius 1\,kpc,
larger and more massive galaxies have a higher projected density
(in M$_{\odot}$\,kpc$^{-2}$). We note that this methodology may lead
to errors for galaxies that are  far from spherical symmetry,
in particular for flat disks. We return to this in \S\,5.3.

Next, the mass within $r=1$\,kpc was calculated by integrating the
3D luminosity profiles. A small
(typically $<10$\,\%) correction to the masses was applied
to take into account that the
total magnitude in the catalog is not identical to the total
magnitude implied by the Sersic fit (see {Taylor} {et~al.} 2010).
This same correction was applied to the total masses used in later
Sections. We also assume that mass follows light. This assumption
is probably reasonable in a relative sense, as there is no evidence
that color gradients (and hence $M/L$ gradients) are a strong
function of redshift (see {Szomoru} {et~al.} 2013). Combining all these
aspects, the core mass is given by
\begin{equation}
M_{\rm 1\,kpc} = \frac{\int_0^{1\,{\rm kpc}} \rho(r) r^2 dr}{\int_0^{\infty}
 \rho(r) r^2 dr}
 \frac{L_{\rm model}}{L_{\rm cat}} M_{\rm cat}.
\end{equation}
Here, $M_{\rm cat}$ is the mass of the galaxy in the 3D-HST catalog,
$L_{\rm cat}$ is the total, aperture-corrected luminosity  of the galaxy
in a particular filter in the 3D-HST catalog, and $L_{\rm model}$ is the
total luminosity implied by the Sersic fit.

Finally, the measurements for $M_{\rm 1\,kpc}$ derived from the $J_{125}$ fit
and from the $H_{160}$ fit were interpolated
so that the final value
corresponds to a rest-frame wavelength of 6000\,\AA\ for $z=1.0-1.7$.
At $z<1$ we use $J_{125}$ and at $z>1.7$ we use $H_{160}$.
We note that this interpolation is not a critical step: the effect
on the derived masses is typically $<5$\,\%, and
using the core mass derived from either $J_{125}$ or $H_{160}$
throughout does not change the results significantly.

The difference between total mass and core mass is illustrated
in Fig.\ \ref{examples.fig}. The figure shows $H_{160}$ images
of three galaxies at $z\sim 1$, along with a graphical representation
of their surface density. The galaxy on the left is a massive elliptical
galaxy with a dense core. Not all massive galaxies have a high central
density: the middle galaxy has a mass that is nearly identical to the
galaxy on the left and a relatively low mass core. The galaxy
on the right is an example of a galaxy whose total mass is only slightly
higher than its core mass; this galaxy resembles the compact quiescent
galaxies that are relatively common at high redshift. The galaxies
on the left and right are included in our $\log M_{\rm 1\,kpc}>10.5$
sample (see \S\,\ref{ndens.sec}), and the galaxy in the middle is not.

We note here that we do not  measure the (deprojected)
light in the central regions directly, but use the best-fitting Sersic
profiles as a proxy for the light at $r<1$\,kpc. The central kpc
covers $\sim 12$ drizzled WFC3 pixels, and we could have used
a direct measurement of the flux within this aperture.
Using the Sersic fits instead takes the effects of the PSF
into account, and enables the deprojection described above.
As described in van der Wel et al.\ (2012, 2014) these fits are
stable even for galaxies with $r_e\sim 1$\,kpc; the total uncertainties
in $n$ and $r_e$ are $<20$\,\% and $<10$\,\% respectively,
for $H_{160}<23$ (see van der Wel et al.\ 2012). 
However, it is
possible that our results are affected by (large) galaxies whose
surface brightness profiles deviate strongly from a
Sersic profile.

\subsection{Low Redshift Samples}

The procedures followed for the UltraVISTA/Z\"urich
galaxies at $0<z<0.5$ and for
the SDSS galaxies at $z\approx 0.06$
were similar to those described above. 
The only difference in procedure
is that the UltraVISTA masses were not corrected for the
difference between total catalog magnitudes and the total fluxes
implied by the GIM2D fit. In Appendix \ref{uvista.sec} we compare
the UltraVISTA masses to the 3D-HST masses for objects that are
in both catalogs. A redshift-dependent offset
was applied to the UltraVISTA masses so they are
consistent with the 3D-HST masses. The origin of this offset
is not understood; we conservatively increase the formal
uncertainty in the UltraVISTA masses by the same amount so that
an offset of zero, or of twice the applied offset, are within the
$1\sigma$ errorbars in all plots.
After applying this offset
the core masses derived from UltraVISTA/Z\"urich are consistent
with those derived from 3D-HST to $0.00 \pm 0.02$ dex
(see \S\,\ref{uvista.sec}).

\begin{figure*}[htbp]
\epsfxsize=16.5cm
\begin{center}
\epsffile[38 322 532 675]{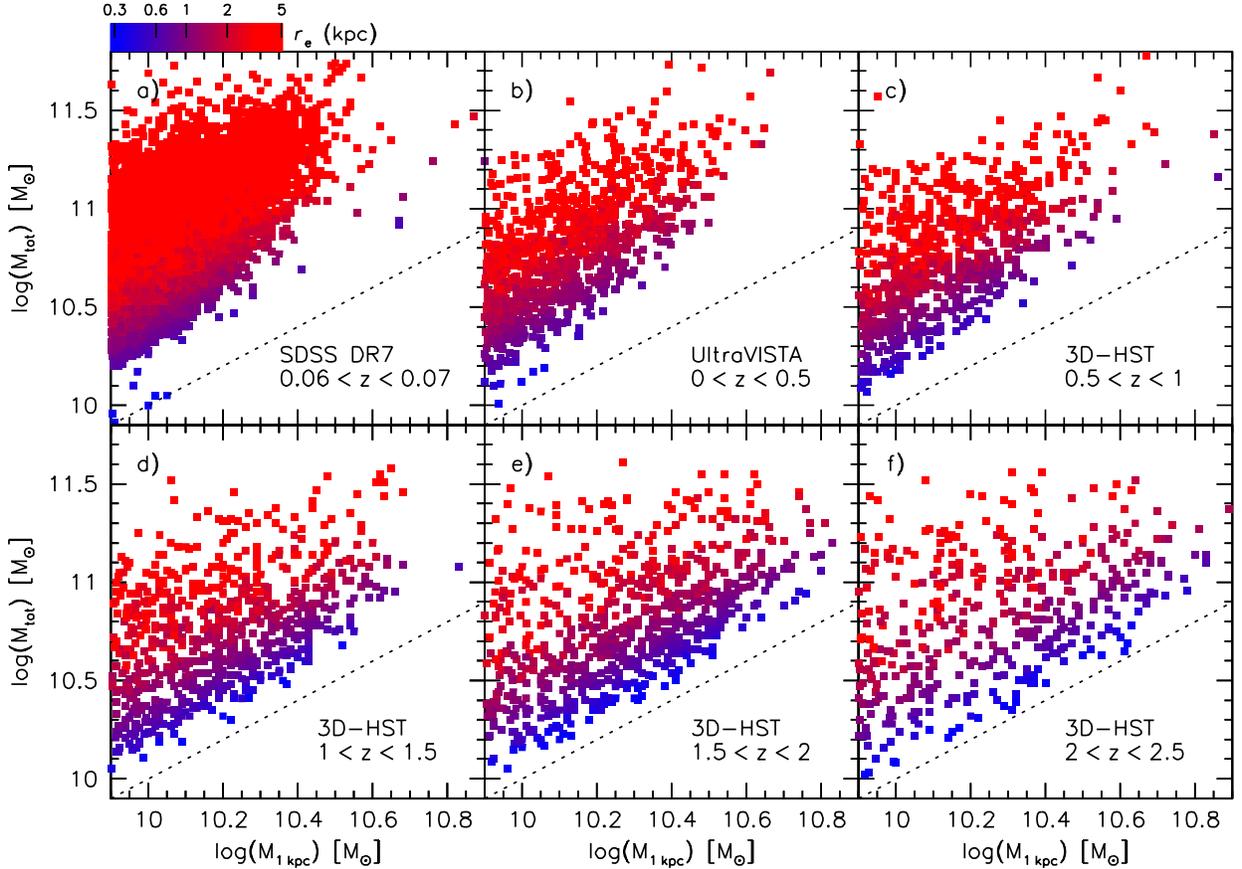}
\end{center}
\caption{\small
Relation between total stellar mass and the stellar mass in a
sphere of radius 1\,kpc. The panels show nearby galaxies from
the Sloan Digital Sky Survey {\em (a)}, galaxies at moderate
redshift from the UltraVISTA and ``Z\"urich'' surveys of the
1.5\,degree$^2$ COSMOS field  {\em (b)}, and distant galaxies
from the 3D-HST survey of the CANDELS fields {\em (c-f)}.
The galaxies are color-coded by their projected, circularized effective
radius. 
At fixed total mass, galaxies have progressively smaller
sizes and higher core masses at higher redshifts.
\label{massmass.fig}}
\end{figure*}

\begin{figure*}[hbtp]
\epsfxsize=16cm
\begin{center}
\epsffile[38 326 532 660]{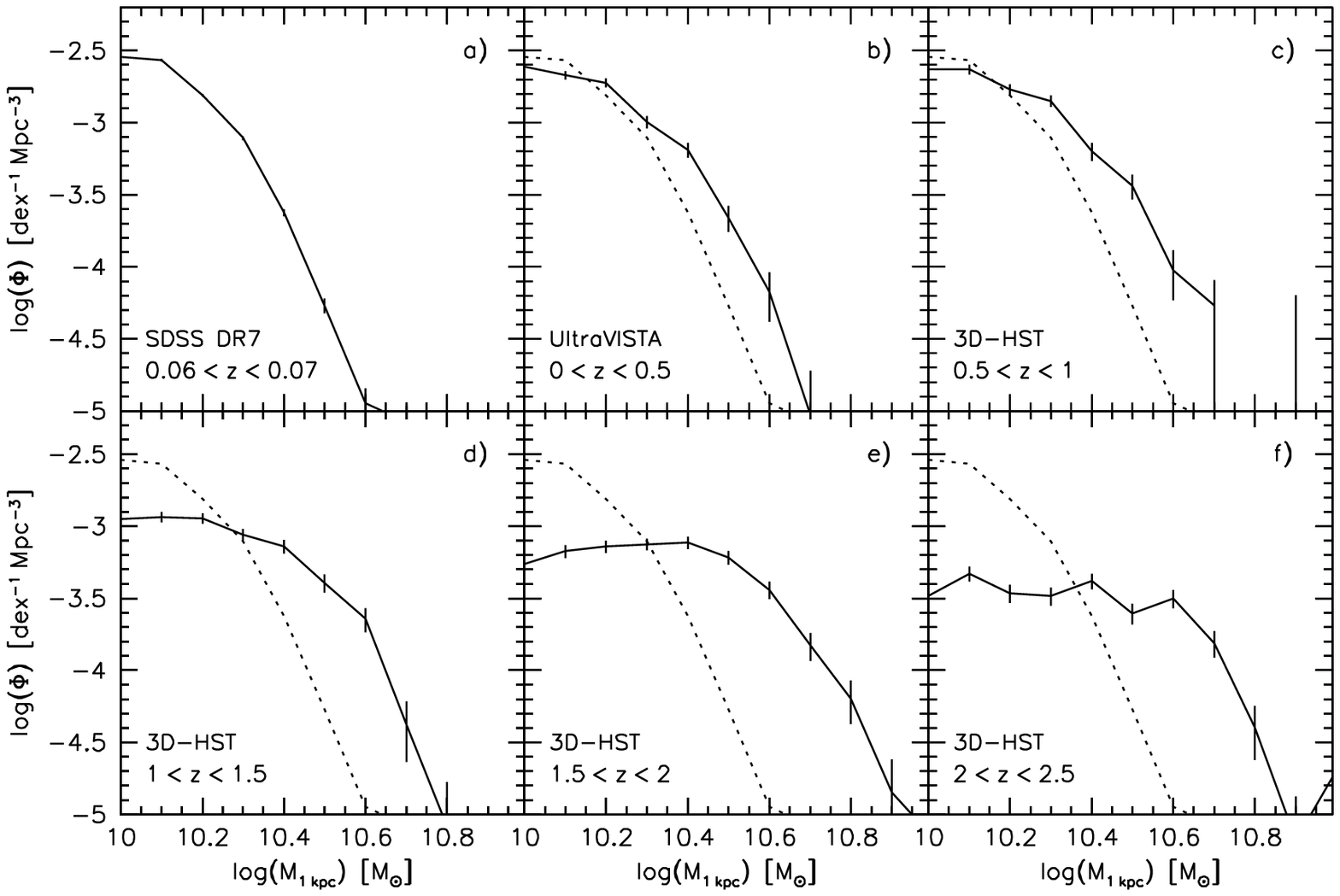}
\end{center}
\caption{\small
Evolution of the core mass function, i.e.,
the number of galaxies as a function of their mass within 1\,kpc.
Panels show data from the SDSS  {\em (a)},
UltraVISTA/Z\"urich {\em (b)}, and 3D-HST {\em (c-f)}. The SDSS function
is repeated in the other panels (broken lines). The shape
of the core
mass function changes with redshift: the number of low mass
cores decreases with redshift, whereas the number of high mass cores
increases.
\label{densfunc.fig}}
\end{figure*}

\subsection{Relation Between Total Mass and Core Mass}

In Fig.\ \ref{massmass.fig} we show the relation between the total mass
of galaxies and the
mass within 1\,kpc in six redshift bins from $z=0$ to $z=2.5$.
All galaxies are shown; there was no selection on star formation rate
or any other property.
The 3D-HST catalog is $90$\,\% -- 95\,\%
complete for masses $>10^{10}$\,\msun\ out to $z=2.5$
({van der Wel} {et~al.} 2014).\footnote{The completeness gradually decreases at
higher redshifts, partially because the 4000\,\AA\ break enters the
observed WFC3 $H_{160}$ filter.}
It is clear from this figure that the central mass is not a fixed
fraction of the total mass: the 68\,\% range in $\log(M_{\rm tot})$ at
fixed $\log(M_{\rm 1\,kpc})$ is approximately 0.5\,dex. A selection
on core mass is therefore distinct from a selection on total mass
(see also, e.g., {Kauffmann} {et~al.} 2003; {Franx} {et~al.} 2008).

Starting with the SDSS sample and going to higher redshifts, the
distribution of points shifts and tilts such that at fixed total mass
galaxies have higher core masses at higher redshifts.
The dashed line shows the line of equality: objects
near this line are so compact that they have close to 100\,\% of their 
mass in the central kpc. At low redshift nearly
all galaxies are far removed
from this regime, particularly at the high mass end. However, at $z>1.5$
the distribution begins to approach this line, with the most
pronounced change at the highest masses.
The galaxies are color-coded by their projected circularized effective
radius.
As is well known, massive galaxies with small sizes are extremely
rare in the nearby Universe ({Taylor} {et~al.} 2009; {Trujillo} {et~al.} 2009).
Mirroring the trend between core mass and total mass, and
consistent with many previous studies (e.g., {Daddi} {et~al.} 2005; {van Dokkum} {et~al.} 2008),
we see that such galaxies are increasingly common
at higher redshifts.

\section{Number Densities}
\label{ndens.sec}

\subsection{The ``Core Mass Function''}

In the remainder of the paper we select objects using
the $x-$axis of Fig.\ \ref{massmass.fig} and 
ask what the properties are of galaxies at fixed core mass rather than
fixed total mass. We first consider the number density of galaxies
as a function of their mass within 1 kpc, the ``core mass function''.
This function is shown in Fig.\ \ref{densfunc.fig} for the same
six redshift intervals as in Fig.\ \ref{massmass.fig}. For
reference, the SDSS core mass function (panel a)
is shown in panels b--f with a dashed line.

The core mass function evolves with redshift in a complex way. Going
from low to high redshift, low mass cores decrease strongly in number
density:  the number of cores with
$\log(M_{\rm 1\,kpc}) = 10$ is
a factor of $\sim 6$ lower at $z=2-2.5$ than it is at $z=0$.
By contrast, the number of high mass cores
apparently {\em in}creases with redshift. The high mass end
of the $z=2-2.5$
core mass function is shifted by $\approx 0.2$ dex in mass compared
to that at $z=0.06$. Because of the steepness of the mass function,
the number of cores with $\log(M_{\rm 1\,kpc})=10.7$ 
increases by more than an order of magnitude from $z=0$
to $z=2.5$.  In Appendix \ref{effect_errors.sec}
we show that random errors do not significantly influence this
result.

This evolution is similar to the evolution of
the (inferred) velocity dispersion function ({Bezanson} {et~al.} 2011):
this function shows a similar differential evolution of high
dispersion and low dispersion galaxies, at least from $z=0.3$
to $z=1.5$. 
This is obviously not a completely independent measurement:
as the velocity dispersions and
central masses are both, to first order,
measures of the compactness of galaxies
they are expected to trace one another.

\subsection{Evolution of the Cumulative Number Density of Massive
Cores}

\begin{figure}[htbp]
\epsfxsize=8.6cm
\epsffile[23 175 575 700]{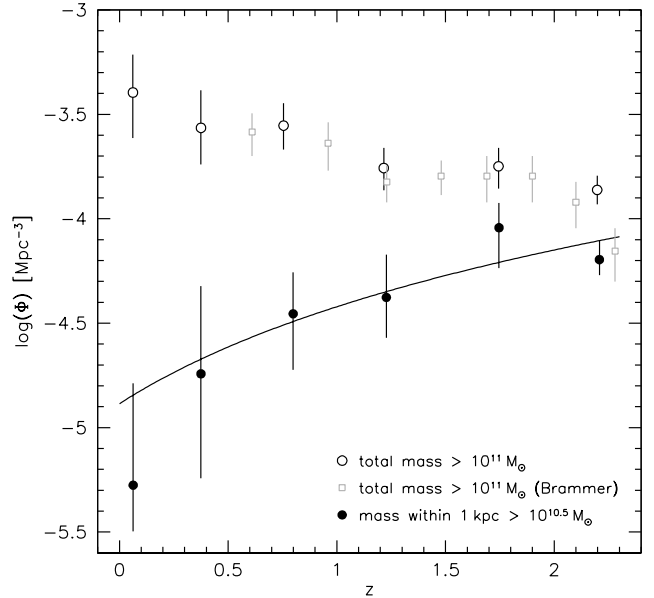}
\caption{\small
Evolution of the number density of galaxies that have
dense cores with $\log (M_{\rm 1\,kpc})
>10.5$ (solid black circles). The number of dense cores was higher in the
past. The line is a fit to the data. Open symbols show the evolution
of massive galaxies with $\log (M_{\rm tot})>11$, from this paper (black
circles) and from Brammer et al.\ (2011) (grey squares). The number density
of massive galaxies decreases with redshift,
and evolves in a very different way than the number density
of dense cores.
\label{dens_evo.fig}}
\end{figure}

We show the evolution of the number density of galaxies with a
central mass $\log (M_{\rm 1\,kpc})>10.5$
in Fig.\ \ref{dens_evo.fig}.
The errorbars reflect the quadratic sum of the Poisson error and a
systematic error, calculated by varying the masses of the galaxies by
$\pm 0.05$\,dex in 3D-HST, $\pm 0.09$\,dex in UltraVISTA, and
$\pm 0.1$\,dex in SDSS. The SDSS error reflects the
systematic uncertainty compared to the 3D-HST masses; in this
paper the 3D-HST ``system'' is taken as the default.
The value of $0.09$\,dex is
the offset that was applied to bring the UltraVISTA data onto
the 3D-HST system (see \S\,\ref{uvista.sec}). 
The evolution can be approximated by the solid line, which 
has the form
\begin{equation}
\label{coreevo.eq}
\log (\Phi) = (-4.9\pm 0.2) + (1.5 \pm 0.5) \log(1+z).
\end{equation}
This increase in the number density of dense
cores with redshift is remarkable, as it
is well established that the mass density of the Universe decreases
rapidly over this same redshift range (e.g., {Dickinson} {et~al.} 2003; {Rudnick} {et~al.} 2003).
Although massive galaxies evolve less rapidly than the overall population
({Marchesini} {et~al.} 2009), even their number density does not  increase with
redshift. We explicitly compare the evolution of dense cores
to the evolution of massive galaxies with $\log(M_{\rm tot})>11$
in Fig.\ \ref{dens_evo.fig}, showing both our data (open circles)
and number densities
of massive galaxies
derived from the NEWFIRM Medium Band Survey by {Brammer} {et~al.} (2011) (open
squares).
At $z\sim 2$, the number density of
massive cores is only about a factor of 2 lower than that of massive galaxies.
At $z\sim 0$, the number density is a factor of $100$ lower.
The striking difference in the evolution
of massive galaxies and massive cores
in Fig.\ \ref{dens_evo.fig} strongly suggests that Eq.\ \ref{coreevo.eq}
is not driven by mass errors, as
errors in the masses should affect the
open circles in the same way as the solid circles.
In Appendix \ref{cosvar.sec} we also show that field-to-field variation
(``cosmic variance'') is not dominating the error budget.

We note that there are
indications of a similar effect
in previous studies. The evolution of the
velocity dispersion function from $z=0.3$ to $z=1.5$ is consistent
with our results ({Bezanson} {et~al.} 2011), and {van de Sande} {et~al.} (2013) find
that, at constant dynamical mass,
the central ($<1$\,kpc) stellar density of quiescent galaxies at $z\sim 2$
is a factor of $\sim 3$ higher than at $z=0$.

\subsection{Effects of Mergers on the Number Density}
\label{mergers.sec}

Setting aside the possibility of errors in the masses or the structural
parameters, there are several plausible explanations
for the observed evolution in Fig.\ \ref{dens_evo.fig}.
We first consider the effects of mergers.
Major mergers can increase the core mass, either through dissipationless
processes (e.g., {Hilz}, {Naab}, \& {Ostriker} 2013) or through merger-induced star
formation ({Solomon}, {Downes}, \&  {Radford} 1992; {Kormendy} \& {Sanders} 1992; {Hopkins} {et~al.} 2008, and many other studies).
The most straightforward
way to {\em de}crease the number
density of galaxies with $\log (M_{\rm 1\,kpc})
>10.5$ is if two galaxies with such dense cores merge with each other.

We determined the number of potential core-core mergers
in the 3D-HST survey,
using the same criteria as {Williams}, {Quadri}, \&  {Franx} (2011) to identify paired
galaxies. In the
entire 3D-HST/CANDELS area we find three galaxy pairs with
a projected distance $d<43$\,kpc, a redshift difference $|z_1-z_2|/(1+z_1)<
0.2$, and $\log (M_{\rm 1\,kpc})>10.5$ for both galaxies.
The pairs are at $z\approx 1.7$, $z\approx 2.0$, and $z\approx 2.3$;
images are shown in Appendix \ref{pair.sec}. The
individual galaxies are
well-separated; no tidal features are detected.
Interestingly all three pairs are red
and in apparent overdensities of other red
objects; we will return to this in a future paper.

The pair fraction is defined as the number of pairs divided by the
number of galaxies in the parent population. The total number of
galaxies with $\log (M_{\rm 1\,kpc})>10.5$ is 273, and we infer that
the pair fraction is $1.1\pm 0.6$\,\%. This is a factor of 
$\sim 5$ lower than the pair fraction in the general population
of massive galaxies ({Bell} {et~al.} 2006; {Bundy} {et~al.} 2009; {Williams} {et~al.} 2011).
Specifically, {Williams} {et~al.} (2011) find a pair fraction of $6\pm 1$\,\%
when requiring that the paired galaxies have masses that are
within a factor of four of one another and
the most massive galaxy has $\log M_{\rm tot}>10.5$.

Turning pair fractions into merger rates is notoriously difficult,
as it depends on the fraction of pairs that is physically bound
and the average time it takes for bound pairs to merge
(e.g., {Bell} {et~al.} 2006; {Kitzbichler} \& {White} 2008).
Nevertheless, it is clear that a pair
fraction of 1\,\% implies a low core-core merger rate. The separation of
the three pairs is $\sim 30$\,kpc. Taking the orbital
time as a lower limit on the merger time scale and assuming
$v\sim 500$\,\kms, we find
that the pairs may merge after $\sim 200$\,Myr. The actual merger
time scale (including a correction for unbound pairs)
is probably significantly larger (see {Kitzbichler} \& {White} 2008).
Using a pair
fraction of 1\,\% we derive an upper limit on the merger rate
of $\sim 5$\,\% per Gyr. We note that if we use a distance limit
of 100\,kpc rather than 43\,kpc we find a higher pair fraction
(of $\sim 2$\,\%) but a similar merger rate, as the orbital time is
longer for pairs with wider separation.
We infer that the 
decline in the number density of galaxies
with massive cores is probably not caused by major mergers between
galaxies with dense cores. 

As the direct effect of core-core merging on the  number density is probably
small, the observed evolution of the core mass function
is most likely a reflection of mass
evolution in the central kpc (or of systematic errors).
As can be seen in Fig.\ \ref{densfunc.fig}  mass evolution of
$-0.15$ to $-0.2$ dex is sufficient to bring the low redshift and high
redshift core mass functions into agreement.
This mass evolution may be the result of a redistribution of
matter following a  merger
(e.g., {El-Zant}, {Kim}, \&  {Kamionkowski} 2004; {Oser} {et~al.} 2012). 
Mergers are thought to be common, particularly at 
redshifts $0<z<1.5$
(e.g., {van Dokkum} 2005; {Naab} {et~al.} 2009; {Newman} {et~al.} 2012,
Bluck et al.\ 2012, and \S\,5).
The central regions of galaxies are thought to
be mostly unaffected by minor mergers
(see, e.g.,
Fig.\ 3 in {Hilz} {et~al.} 2013). This is also suggested by
the results of {Weinmann} {et~al.} (2013), who
have shown that the number density of dark matter halos with
fixed high central density is constant to within $\sim 0.2$ dex
at $0<z<4$. We stress, however, that
the quantitative effects of  mergers on the mass profile
within 1\,kpc are not well known.

\subsection{Effects of Stellar Winds}
\label{massloss.sec}

As is well known ``negative'' mass evolution
is expected (and inevitable)
in an isolated stellar population:
the mass locked up in stars and stellar remnants
decreases with time due to supernova explosions and stellar winds.
The amount of mass loss is a strong function of the mass
and surface gravity of the
stars that are present in the population, and therefore a strong function
of both the age  and the IMF.

Previous
studies have considered the effects of mass loss
on the evolution of the mass-size relation,
particularly in the context of the size
growth of quiescent galaxies ({Damjanov} {et~al.} 2009).
As mass loss leads to adiabatic expansion, passively
evolving galaxies become both less
massive and larger with time; however, these effects are small
compared to the observed evolution of the sizes and masses
of quiescent galaxies
({Damjanov} {et~al.} 2009; {Bezanson} {et~al.} 2009; {Ragone-Figueroa} \& {Granato} 2011).

Here we are concerned with a slightly different question, namely
the effect on the mass contained within a sphere of 1 kpc. The
mass within a fixed radius is affected by stellar mass loss in
two ways. The first, direct, effect is that the stellar mass
measurements of galaxies at later times will be lower, as they only
include living stars and stellar remnants and not the mass that is
lost during stellar evolution. The second effect is that the matter
in galaxies can be redistributed as a result of adiabatic expansion.
This effect is only important 
if the material in
the winds mixes with the hot halo gas, which in turn depends on the
details of the interaction of the wind material with the ambient gas
(e.g., {Bregman} \& {Parriott} 2009, Conroy et al.\ 2014). We show in
Appendix \ref{lossaper.sec} that the
total effect of stellar mass loss is
\begin{equation}
\label{adia.eq}
(M'/M)_{\rm 1\,kpc} 
   \sim  (M'/M)_{\rm tot}^{1.8},
\end{equation}
with $(M'/M)_{\rm tot}$ the change in the total mass and
$(M'/M)_{\rm 1\,kpc}$ the change in mass in the central kpc.
The relation in Eq.\ \ref{adia.eq} assumes that
all the material escapes and there are no other sources of mass
loss than stellar evolution. If no material escapes, Eq.\ \ref{adia.eq}
is simply $(M'/M)_{\rm 1\,kpc} \sim (M'/M_{\rm tot}$.

\begin{figure}[htbp]
\epsfxsize=8.2cm
\epsffile[35 180 308 657]{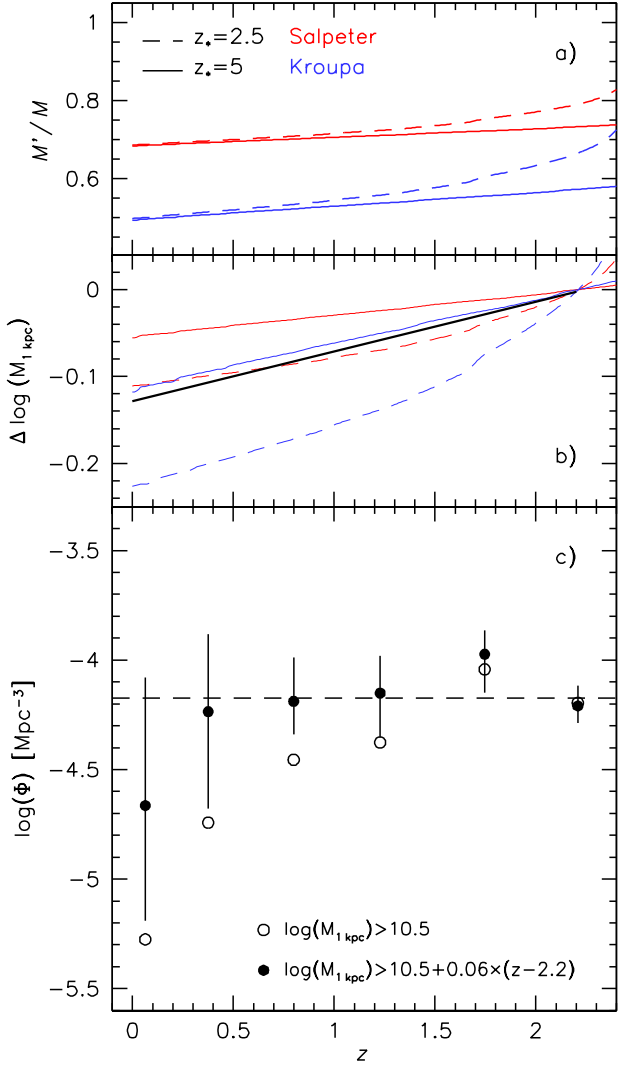}
\caption{\small
{\em a)} Stellar mass evolution of a passively evolving stellar
population. The mass decreases with time, mostly due to winds from
AGB and post-AGB stars. Mass loss is largest for young stellar
populations and bottom-light IMFs. {\em b)}
Evolution of the
mass within 1 kpc.
The evolution is stronger than in panel a) due to the
effects of adiabatic expansion, under the assumption that the stellar
ejecta turbulently mix with the ambient hot gas.
The black line is a linear fit to the
$z_*=2.5$ Salpeter model, of the form $\Delta \log (M_{\rm 1\,kpc})
= 0.06z$. {\em c)} The number density of dense
cores, after correcting the core masses for stellar mass loss
according to the black line in panel b). The data are consistent
with a constant number density of $\sim 7\times 10^{-5}$\,Mpc$^{-3}$
(dashed line).
\label{massloss.fig}}
\end{figure}

The change in the total mass due to stellar evolution can be
estimated using stellar population synthesis models. The main
uncertainty is the age of the stars, with the added complexity
that the ages in the central kpc can be different from the ages
at larger radii. We consider two mean formation redshifts of the
stars, $z_*=2.5$ and $z_*=5$, and two IMFs, a Milky-Way
like {Kroupa} (2001) IMF and a bottom-heavy {Salpeter} (1955) IMF.
The top panel of Fig.\ \ref{massloss.fig} shows the mass evolution
of a passively evolving stellar population for the four model
combinations, determined using the {Bruzual} \& {Charlot} (2003) stellar
population synthesis model. At late times
the mass loss is approximately 30\,\% for a Salpeter IMF and
50\,\% for a Kroupa IMF. The relative mass loss from $z=2.5$ to
$z=0$ depends on the combination of the IMF and the
formation redshift of the stars. In Fig.\ \ref{massloss.fig}b
we show the effect on the mass within 1\,kpc, relative to $z=2.2$
(our highest redshift bin). The minimum effect of mass loss
is $\approx 0.06$ dex (for $z_*=5$ and a bottom-heavy IMF)
and the maximum effect is $\approx 0.22$ dex (for $z_*=2.5$ and
a bottom-light IMF).

In Fig.\ \ref{massloss.fig}c we show the effect of mass loss on
the evolution of the number density of dense cores. We adapted the
selection of galaxies, so that we select cores of lower mass at lower redshift.
This redshift dependence is given
by the relation $\Delta \log(M_{\rm 1\,kpc})=0.06\times z$
(solid black line in Fig.\ \ref{massloss.fig}b), which is
a linear fit to the mass loss expected for
a {Salpeter} (1955) IMF with $z_*=2.5$. This is also a good fit to
the mass loss for a {Kroupa} (2001) IMF with $z_*=5$,
but it is lower than the mass loss expected for a {Kroupa} (2001)
IMF with $z_*=2.5$ (see Fig.\ \ref{massloss.fig}b).
The solid
points in Fig.\ \ref{massloss.fig}c show the number density of
galaxies that have
$\log(M_{\rm 1\,kpc})>10.5+0.06\times (z-2.2)$.
The errorbars reflect the systematic uncertainties in the
masses, as before. We find that the data are now consistent with
a constant number density since
$z=2.5$, within the uncertainties.
The weighted mean number density is $\Phi \approx 7\times 10^{-5}$\,Mpc$^{-3}$.

\begin{figure*}[htbp]
\epsfxsize=16cm
\begin{center}
\epsffile[50 426 530 660]{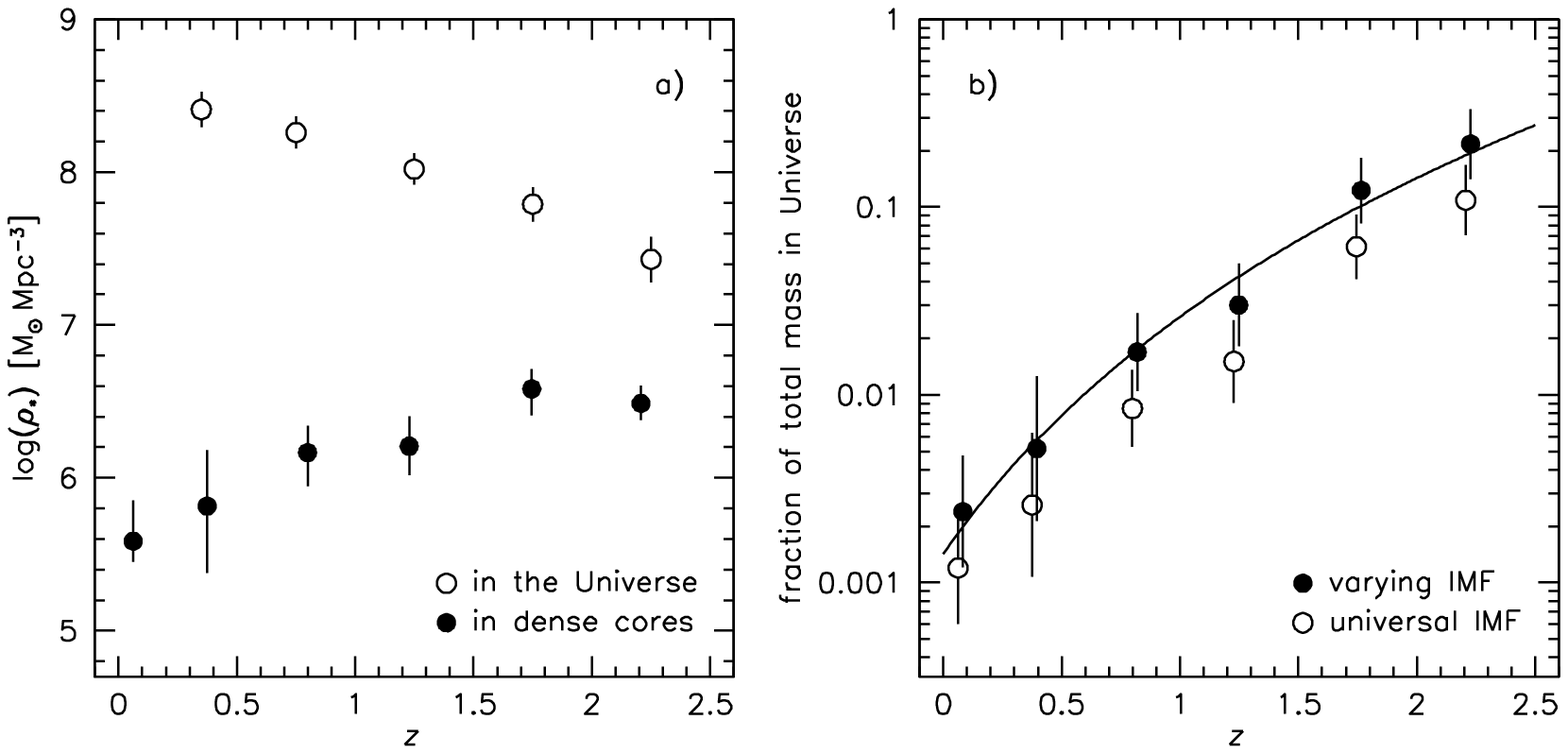}
\end{center}
\caption{\small
{\em a)} Total stellar mass contained in cores with
$\log(M_{\rm 1\,kpc})>10.5$, as a function of redshift (solid circles).
Open circles show
the total stellar mass density of the Universe, as determined by
{Muzzin} {et~al.} (2013a). A {Kroupa} (2001) IMF was assumed, for
the cores and for the Universe. {\em b)}
Fraction of the total stellar mass density of the Universe that
is contained in cores with 
$\log(M_{\rm 1\,kpc})>10.5$. Open symbols assume a universal
IMF, solid symbols (slightly offset for clarity) assume that the
stellar mass function
in dense cores has a factor of two more mass than that in
the rest of the Universe, as might be expected
if the IMF in star-forming cores
was bottom-heavy ({Conroy} \& {van Dokkum} 2012).
Stars in dense cores comprise 10\,\% -- 20\,\% of the stellar
mass density at $z>2$, and reflect an important mode of
star- and galaxy formation in the early Universe.
\label{fractions.fig}}
\end{figure*}

We conclude that the mass evolution in the central kpc can be
explained with a passively evolving population formed at $z>2.5$,
under the assumption that 100\,\% of the mass lost during stellar
evolution mixes with the hot halo gas. In reality, the central
mass decrease probably reflects a combination of the effects of mergers
and mass loss, and it will be difficult to disentangle these effects.

\subsection{Other Explanations}
\label{other.sec}

We briefly consider several
other explanations for the apparent negative
evolution of the number density of dense cores.
One possibility is that the structure of galaxies changes in such
a way that the deprojected core masses change systematically with
redshift. Specifically, in the deprojection we use the circularized effective
radius and if the mean flattening of galaxies changes with
redshift this may lead to artifical offsets between the high
redshift and low redshift data. In Appendix \ref{major.sec} 
we show the core mass function derived using the major axis effective
radius rather than the circularized effective radius. The function
is offset to lower masses, as expected, but there is no significant
redshift-dependent effect.

Another interesting possibility is that the mass in the cores decreases
due to scouring by binary supermassive black holes. The presence of flat
density profiles in the centers of massive, slowly
rotating elliptical galaxies\footnote{Also termed ``cores''; see Footnote 12.}
({Faber} {et~al.} 1997) has been attributed to the
ejection of stars by a binary black
hole (see, e.g., {Begelman}, {Blandford}, \&  {Rees} 1980; {Milosavljevi{\'c}} \& {Merritt} 2001; {Kormendy} \& {Bender} 2009; {Hopkins} \& {Hernquist} 2010).
The total amount of mass that is ejected is of order $M_{\rm ej}
\sim M_{\bullet}$ (e.g., {Milosavljevi{\'c}} \& {Merritt} 2001). Our adopted core mass
limit of $3\times 10^{10}$\,\msun\ within 1\,kpc corresponds to a
velocity dispersion of $\sigma \sim 280$\,\kms, which in
turn corresponds to a black hole mass of $M_{\bullet}
\sim 5\times 10^8$\,\msun\ ({G{\"u}ltekin} {et~al.} 2009).
The ratio of the black hole mass to the stellar mass within 1\,kpc is
therefore $\sim 2$\,\%. Even
taking adiabatic expansion into account, we infer that
the stellar mass within 1\,kpc is only
reduced by 0.01 -- 0.02 dex due to black hole scouring, unless
$M_{\rm ej}\gg M_{\bullet}$ (see {Hopkins} \& {Hernquist} 2010, for a recent discussion
on this topic).

Perhaps the most
important alternative explanation is systematic error. A
redshift-dependent error in the stellar masses of $0.05 - 0.1$
dex per unit redshift could fully explain the evolution.
Combined with the (inevitable) effects of mass loss, the
(small) effects of black hole scouring, and the (uncertain)
effects of merging, we conclude that 
the uncertainties 
allow a factor of $\sim 2$ evolution in the number density, in either
direction.

\subsection{Contribution of Stars in Dense Cores to the Stellar
Mass Density of the Universe}
\label{contrib.sec}

As discussed in the Introduction,
the dense cores studied in this paper are extreme environments
by local Universe standards. They contain only a small fraction
of the total stellar mass in the Universe, and as we
showed in Fig.\ \ref{dens_evo.fig} the number
density of galaxies with $\log M_{\rm 1\,kpc}>10.5$ is a factor
of $\sim 100$ lower than that of galaxies with $\log M_{\rm tot}>11$.

However, the fact that their number density and mass
does not increase with time, and probably even decreases, means
dense cores are an increasingly important environment at higher redshift.
Going back in time from the present to $z\sim 2.5$,
we see galaxies such as the Milky Way ``lose'' $\sim 90$\,\% of their
stars ({van Dokkum} {et~al.} 2013) whereas the dense cores become ever
more prominent and striking:
their mass increases and the galaxies in which
they are embedded today are stripped away (see \S\,\ref{buildup.sec}).

Figure \ref{fractions.fig}a shows the integrated stellar mass in cores
of mass $\log(M_{\rm 1\,kpc})>10.5$ as a function of redshift.
This is {\em not} the total stellar mass of all galaxies hosting
such cores: only stars within the central 1\,kpc are counted.
The stellar mass density rises steeply with redshift, reflecting the
increase in the number density shown in Fig.\ \ref{dens_evo.fig}.
Figure \ref{fractions.fig}a also shows the stellar mass density
of the Universe,
taken from Table 2 of {Muzzin} {et~al.} (2013a). Muzzin et al.\ calculated
these mass densities
by integrating the best-fitting Schechter functions
down to $10^8$\,\msun. We added 0.1\,dex in quadrature to the uncertainties
listed in {Muzzin} {et~al.} (2013a)
to account for possible systematic differences between 3D-HST
and UltraVISTA (see Appendix A). The stellar mass density of the Universe
decreases with redshift, and the stars living in dense cores make up a
rapidly increasing fraction of the total stellar mass density.

In Fig.\ \ref{fractions.fig}b we show this fraction explicitly. It
rises from $\sim 0.1$\,\% at $z=0$ to $\sim 10$\,\% at $z=2$. The fraction
is even higher if the IMF is bottom-heavy in dense regions, as has been
inferred from studies of absorption lines and the masses of nearby
galaxies (e.g., {van Dokkum} \& {Conroy} 2010; {Treu} {et~al.} 2010; {Cappellari} {et~al.} 2012; {Conroy} \& {van Dokkum} 2012).
Somewhat dependent on the detailed functional form of the IMF,
{Conroy} \& {van Dokkum} (2012) find that
galaxies with high velocity dispersions 
have stellar masses that are a factor of $\sim 2$
higher than would be inferred from a {Kroupa} (2001) IMF.
The solid points in Fig.\ \ref{fractions.fig}b indicate the effect of
such IMF variation. The line is a fit to the solid points of the form
\begin{equation}
f = 1.4\times 10^{-3} \times (1+z)^{4.2},
\end{equation}
with $f$ the fraction of the total stellar mass density in the Universe
that is locked up in dense cores.
The implication is that the stars in dense
cores may constitute $\sim 20$\,\% of the total stellar mass density at
$z=2.2$. Phrased differently, $\sim 20$\,\% of all
the stars formed at $z\gtrsim 2.5$ may have ended up in dense cores.

\section{Build-Up of Galaxies Around Dense Cores}
\label{buildup.sec}
\subsection{Evolution of Total Mass and Effective Radius}

We now turn to the properties of the galaxies in which the dense
cores reside.
The fact that the number density of galaxies with dense cores
is relatively stable with time has an important implication:
it means that we can plausibly identify the descendants of
high redshift galaxies with dense cores. In the general
population of massive galaxies
the number density at fixed mass increases
with time, and as different galaxies can have different growth rates
there is an inherent uncertainty in linking
progenitors and descendants (e.g., {Newman} {et~al.} 2012; {Leja}, {van Dokkum}, \& {Franx} 2013).
In contrast, galaxies with dense cores are consistent
with having
a passive stellar population in their central 1\,kpc, enabling,
in principal, a one-to-one matching of galaxies across cosmic time.

In practice processes such as mergers, stellar winds, occasional
star formation, and black hole scouring make this comparison
less straightforward (see \S\,4.3 -- 4.5).
Here we take mass loss due to stellar winds
into account but ignore all other
effects, including possible systematic errors.
We use the black solid line in Fig.\ \ref{massloss.fig}b to
parameterize mass loss:
\begin{equation}
\label{massevo.eq}
\log({\rm M}_{\rm 1\,kpc}^{\rm evo})\,[{\rm M_{\odot}}] = 10.37 + 0.06z.
\end{equation}
As discussed in \S\,\ref{massloss.sec}
this parameterization is a linear fit to the mass loss expected for
a {Salpeter} (1955) IMF with $z_*=2.5$.
We measure the build-up of massive
galaxies around their dense cores
by selecting galaxies in a narrow mass bin centered on this evolving
mass.

\begin{figure*}[htbp]
\epsfxsize=17.5cm
\begin{center}
\epsffile[28 425 530 658]{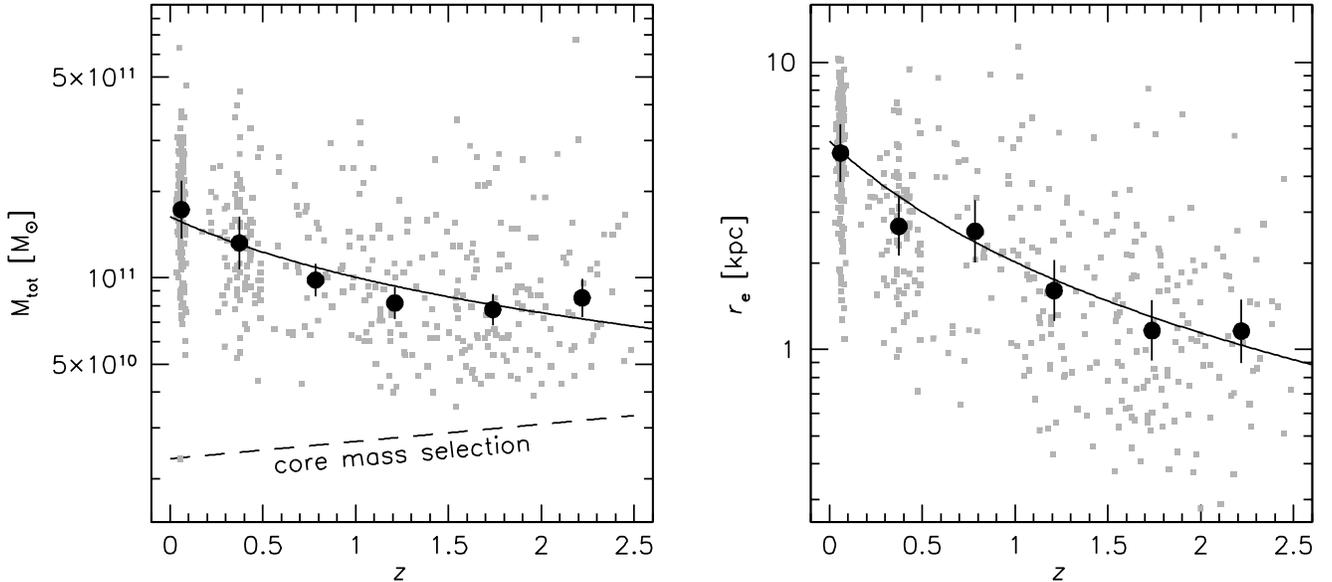}
\end{center}
\caption{\small
Left panel: Total stellar mass of galaxies that have
a core mass $\log(M_{\rm 1\,kpc})=(10.37+0.06z)\pm 0.05$. 
Black symbols show the means for SDSS ($z=0.06$),
UltraVISTA/Z\"urich ($z=0-0.5$), and
3D-HST ($z=0.5-2.5$).
The mean mass evolves as 
$M_{\rm tot} \propto (1+z)^{-0.7}$. Dense cores make up $\sim 50$\,\% of
the total mass at $z=2.5$ and $\sim 15$\,\% at $z=0$.
Right panel: Effective radii of galaxies
with dense cores. The effective radius evolves as
$r_e \propto (1+z)^{-1.4}$.
\label{evo.fig}}
\end{figure*}

Figure \ref{evo.fig}(a) shows the total masses of galaxies with
a core mass $\log(M_{\rm 1\,kpc}) = \log({\rm M}_{\rm 1\,kpc}^{\rm evo})
\pm 0.05$. The total masses increase with time, despite the fact that we
selected the galaxies to have a central mass that {\em de}creases with time.
The total mass evolves as
\begin{equation}
\label{mtotevo.eq}
\log (M_{\rm tot})\,[{\rm M}_{\odot}]
= (11.21 \pm 0.04) - (0.70 \pm 0.08) \log(1+z),
\end{equation}
with the errors determined from bootstrap resampling. This fit
is indicated by the solid line in Fig.\ \ref{evo.fig}(a).
The evolution is slightly faster if we correct for stellar mass loss.
As $(M'/M)_{\rm tot} \sim (M'/M_{\rm 1\,kpc})^{0.6}$ this
correction $\Delta
\log(M_{\rm tot}) \approx 0.6 \times 0.06z \approx 0.04z$.
The mass evolution is then $\log (M_{\rm tot,cor}) \approx 11.30 - 0.87
\log(1+z)$. 
Going to higher redshifts,
dense cores make up an increasing fraction of the total mass of the
galaxies that they are part of: at $z=0$ their contribution is
$\approx 15$\,\% and at $z=2.5$ it is $\approx 50$\,\%.

The evolution of the projected, circularized half-light radius
is shown in panel (b) of Fig.\ \ref{evo.fig}.
There is strong evolution,
such that galaxies with dense cores are more compact at high
redshift than at low redshift.
The evolution can be described by
\begin{equation}
\log (r_e)\,[{\rm kpc}] = (0.73\pm 0.06)
 - (1.40 \pm 0.11) \log(1+z).
\end{equation}
The increase in the effective radius is partly caused by adiabatic
expansion as a result of mass loss. Using Eq.\ \ref{expand.eq}
we estimate that the expansion $\Delta \log (r_e) \sim 0.04z$,
and the corrected evolution is $\log (r_e) \approx 0.64 -1.23
\log(1+z)$. The effects of adiabatic expansion are not
negligible but small compared to the observed evolution, as
previously discussed by {Damjanov} {et~al.} (2009) in the context of
the evolution of the mass -- size relation.
Note that these corrections assume
that mass loss is separable from the processes
that cause the increase in the total mass and the effective 
radius.

We conclude that the average mass and size of galaxies with high core
masses evolves with redshift. At high redshift these galaxies are
typically compact, with half-light radii of $\sim 1$\,kpc and
about half of the total mass contained in the core. At $z=0$
they are embedded in a large envelope of stars and
have effective radii of $\sim 5$\,kpc.
These conclusions apply to the population of galaxies that have a massive
core, and (as discussed at the
beginning of this Section) probably
also describe the mean evolution of individual
galaxies. The cores can then be
interpreted as ``seeds'' around which massive galaxies assemble
over time. From $z=2.5$ to $z=0$, the masses of galaxies
with dense cores increase by a factor
of $\sim 2.4$ and their sizes increase
by a factor of $\sim 6$. These results
are similar to those derived using number density matching
techniques (e.g., {van Dokkum} {et~al.} 2010; {Patel} {et~al.} 2013).

\subsection{Stellar Populations}
\label{spectra.sec}

\begin{figure*}[htbp]
\epsfxsize=17.5cm
\begin{center}
\epsffile[54 434 550 715]{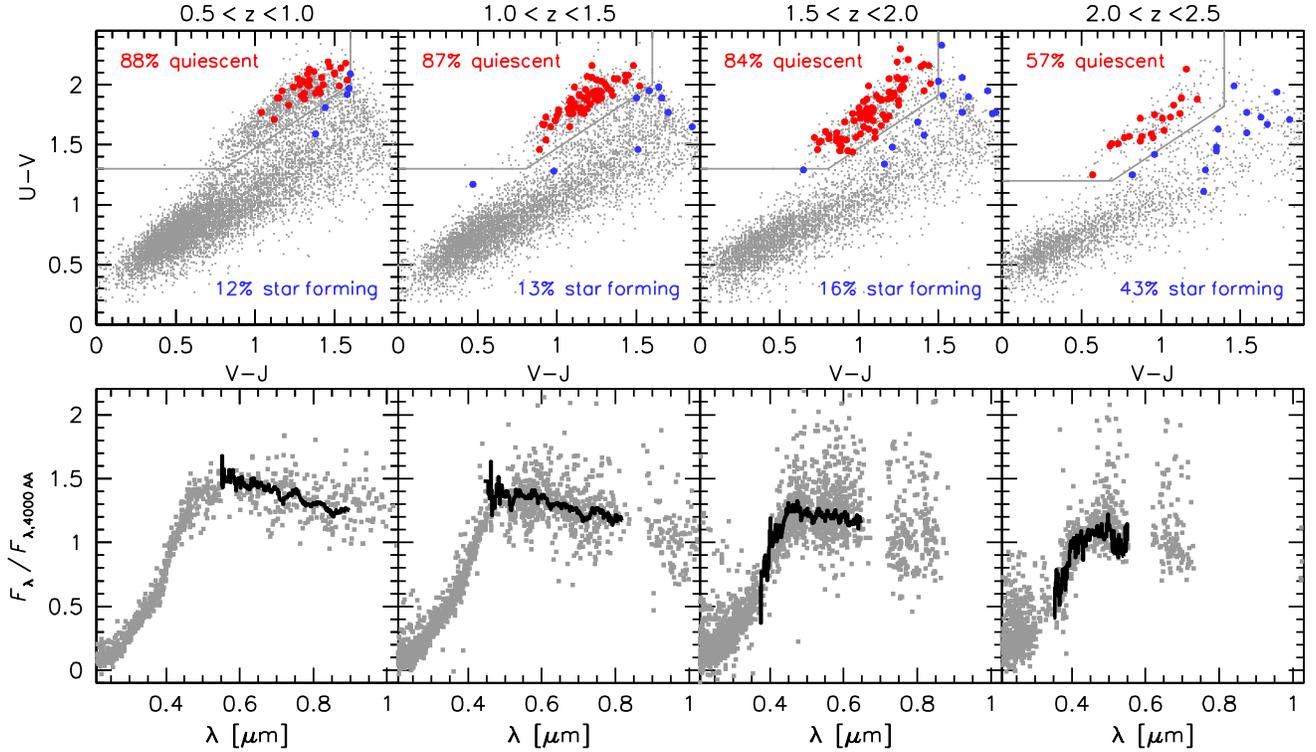}
\end{center}
\caption{\small
{\em Top panels:} Location of galaxies with $\log
(M_{\rm 1\,kpc}) = (10.37 + 0.06z)\pm 0.05$ in
the rest-frame $UVJ$ plane.
Out to $z=2$ more than 80\,\% of the galaxies have colors indicating
evolved stellar populations. The galaxies are among the reddest
quiescent galaxies in the local Universe. At higher redshifts there is
a significant population of star forming galaxies with dense
cores. {\em Bottom panels:}
Stacked SEDs (grey points) and WFC3 grism spectra (black lines)
of the galaxies.
The galaxies have strong Balmer/4000\,\AA\ breaks, confirming that
their light is dominated by cool stars.
\label{stacks.fig}}
\end{figure*}

Galaxies with dense cores have low star formation rates compared
to other galaxies with the same total mass and redshift.
In the top panels of Fig.\ \ref{stacks.fig} we show the location of the
galaxies with dense cores in the $UVJ$ diagram
({Labb{\'e}} {et~al.} 2005; {Wuyts} {et~al.} 2007; {Whitaker} {et~al.} 2011). At fixed rest-frame
$U-V$ color
quiescent galaxies can be separated from dust-reddened star-forming
galaxies by their $V-J$ color (see {Wuyts} {et~al.} 2007). 
The formal errors in Fig.\ \ref{stacks.fig} are small, although we note
that 1) the colors of galaxies near the extremes of the distribution are
influenced by the color range of the EAZY templates, and 2) the location
of the boundary between star-forming and quiescent galaxies is somewhat
arbitrary.
The colors
of the galaxies suggest that their rest-frame optical light
is dominated by relatively evolved stars, and not by
a  (reddened) young stellar population. The fraction of quiescent
galaxies is $>80$\,\% out to $z=2$ and 57\,\% at $2<z<2.5$.

The dominance of relatively cool stars in these galaxies
is demonstrated directly in
the bottom panels of Fig.\ \ref{stacks.fig}, where we show stacked
rest-frame SEDs and WFC3 grism spectra of the galaxies. The spectral
stacks
were created by de-redshifting the observed data, normalizing in
a fixed rest-frame wavelength interval from 4100\,\AA\ -- 4150\,\AA,
and averaging. The SED stacks were created in the same way, except
that we plot the individual data points rather than an average in
wavelength bins. In all
redshift bins the Balmer/4000\,\AA\ break is clearly detected in the
stacked SED and/or the stacked spectrum. The stacked spectra also
unambiguously
demonstrate that the rest-frame optical emission (and therefore the
size and mass measurements) is not greatly influenced
by redshift errors or AGN emission (see also {Whitaker} {et~al.} 2013).

\begin{figure}[htbp]
\epsfxsize=8.2cm
\epsffile[43 170 330 570]{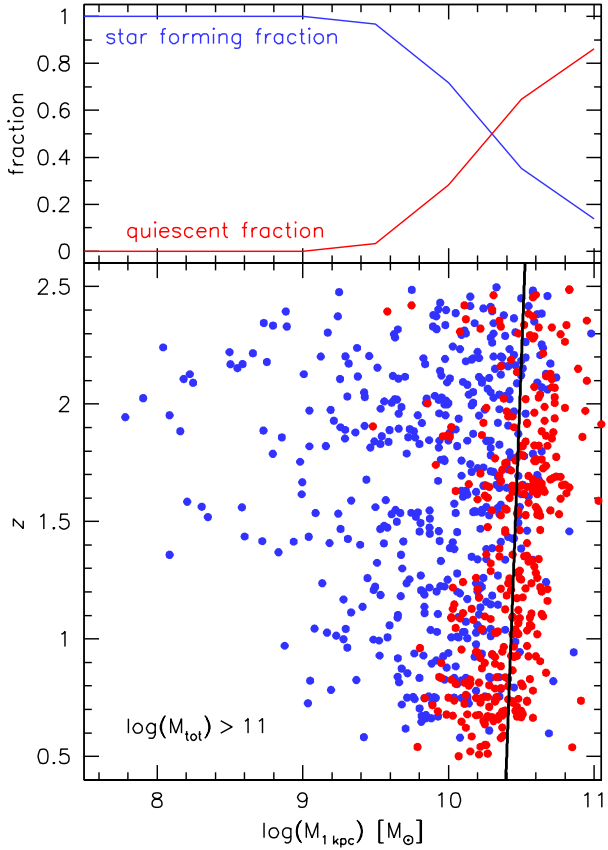}
\caption{\small
Relation between star formation and core mass, for galaxies with
total mass $M_{\rm tot}>10^{11}$\,\msun.
Blue points are galaxies that fall in the star-forming part of the
$UVJ$ diagram; red points are galaxies that fall in the quiescent
part. The black line shows our adopted core mass limit
(Eq.\ \ref{massevo.eq}). The top panel shows the fractions of
quiescent and star forming galaxies as a function of core mass.
At fixed total mass there is a clear relation between core mass
and star formation, and
a high core mass seems to be required to stop star formation.
\label{massfrac.fig}}
\end{figure}

In Fig.\ \ref{massfrac.fig} we show the correlation between core mass
and quiescence, for all galaxies with total masses
$>10^{11}$\,\msun. The main panel of
Fig.\ \ref{massfrac.fig} shows the core
mass as a function of redshift,
with galaxies color-coded by their location in the $UVJ$ diagram.
It is striking how well core mass correlates with star formation rate
(or, more precisely, with the location in the $UVJ$ diagram).
The top  panel shows the fraction of
quiescent and star forming galaxies as a function of core mass.
Essentially
all galaxies that lack a dense core are forming stars, and the majority
of galaxies that have a dense core is quiescent.
This result is consistent
with the fact that high stellar density is an excellent predictor of
quiescence (e.g., {Franx} {et~al.} 2008).

\subsection{Star Forming Galaxies with Dense Cores}

Although most galaxies with dense cores are quiescent, there are
some that fall in the star forming region
of the $UVJ$ diagram. Many of these are
red in both $U-V$ and $V-J$, indicating significant absorption
by dust ({Labb{\'e}} {et~al.} 2005; {Wuyts} {et~al.} 2007; {Marchesini} {et~al.} 2014).
The fraction of star forming
galaxies with dense cores is as high as $\sim 40$\,\% at $z>2$
(top right panel of Fig.\
\ref{stacks.fig}). This is
consistent with many previous studies that have shown that massive
star-forming galaxies exist at these redshifts
(e.g., {F{\"o}rster Schreiber} {et~al.} 2006; {Kriek} {et~al.} 2009a; {F{\"o}rster Schreiber} {et~al.} 2011; {Williams} {et~al.} 2010; {van Dokkum} {et~al.} 2010; {Brammer} {et~al.} 2011; {Marchesini} {et~al.} 2014), and
with the fact that some of these galaxies have small sizes
(e.g., {Patel} {et~al.} 2013; {Barro} {et~al.} 2013).

The presence of
star-forming galaxies with a high central density
does not necessarily
mean that we are witnessing the build-up of the dense cores themselves.
Significant star formation in the cores would lead to an increase
in the core mass, and an increase in the number density of galaxies
with dense cores with time -- which may be difficult to reconcile
with Figures \ref{dens_evo.fig} and \ref{massloss.fig}.

\begin{figure}[htbp]
\epsfxsize=8.3cm
\epsffile[18 157 530 661]{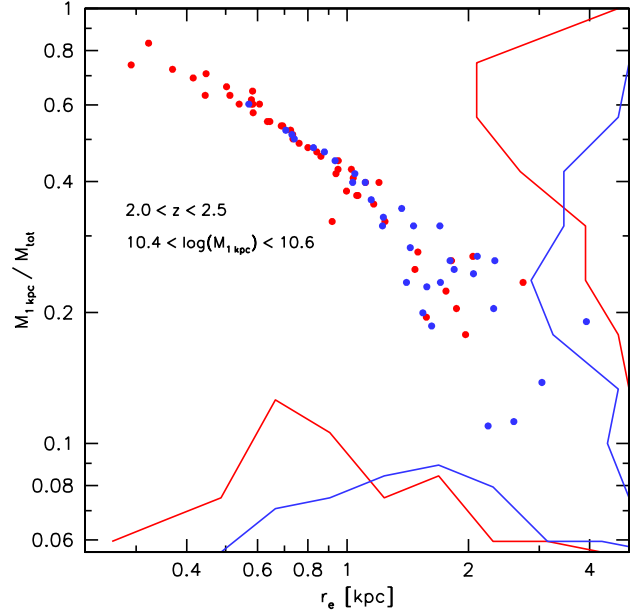}
\caption{\small
The contribution of the central mass to the total galaxy mass
as a function of the projected half-light radius, for $2<z<2.5$
and $\log(M_{\rm 1\,kpc}) = 10.5$.  Red symbols
are quiescent galaxies according to the UVJ diagram; blue
symbols are star-forming galaxies. Star-forming galaxies are
larger than quiescent galaxies at fixed central mass and redshift,
and their cores constitute a smaller fraction of the total
galaxy mass.
\label{fracsize.fig}}
\end{figure}

Instead, many
of these star forming galaxies could be building mass outside
of their centers.
As shown in Fig.\ \ref{fracsize.fig}, $z>2$ star-forming galaxies with
dense cores have a median size that is a factor of $\sim 2$ larger
than that of quiescent galaxies with the same central mass and
redshift, which means they are probably not their direct progenitors.
Furthermore, for the star-forming galaxies the central 1\,kpc
contributes only $\sim 25$\,\% of
the total mass, which means
it is plausible that these galaxies are
building up their outer parts rather than their centers.
This is
qualitatively consistent with other studies of distant galaxies
(e.g., {Labb{\'e}} {et~al.} 2003; {Nelson} {et~al.} 2012, 2013; {Wuyts} {et~al.} 2013; {Genzel} {et~al.} 2013)
and the properties of spiral galaxies in the local Universe.

\subsection{The Fate of Compact Quiescent Galaxies at $z\sim 2$}

The strong mass and size evolution we find for galaxies with dense cores
has implications for the evolution of the general population of
quiescent galaxies. Many studies have shown that quiescent galaxies
with masses of $\sim 10^{11}$\,\msun\ are very compact at $z\gtrsim 2$,
having half-light radii of $\sim 1$\,kpc
(e.g., {van Dokkum} {et~al.} 2008; {Cimatti} {et~al.} 2008; {van der Wel} {et~al.} 2014).
Most of these compact quiescent
galaxies should have a dense core according to our definition, as
galaxies with a projected half-light radius of $\sim 1$\,kpc have
$M_{\rm 1\,kpc} \sim 0.5 M_{\rm tot}$. 

The median circularized effective
radius of the 31 quiescent galaxies with
$10.9 < \log M_{\rm tot}< 11.1$ and $2<z<2.5$ is $r_e = 1.2$\,kpc, consistent
with earlier results. The median core mass of these galaxies
$\log M_{\rm 1\,kpc} = 10.6$, with an rms scatter of 0.2 dex.
Conversely, the median effective radius of galaxies with
$10.4 < \log M_{\rm 1\,kpc} < 10.6$ and $2<z<2.5$ is $r_e = 1.0$\,kpc,
their median total mass $\log M_{\rm tot} =10.9$,
and 60\,\% of these galaxies are quiescent. We conclude that,
at $z=2-2.5$, there
is substantial overlap
between the population of massive quiescent
galaxies and the population of galaxies with dense cores (see Fig.\
\ref{massmass.fig}).

As we have shown that 1) the number density of galaxies with dense
cores does not increase with time; 2) the half-light radii and total masses of
galaxies with dense cores grow with time; and 3) the population of galaxies
with dense cores overlaps substantially with the population of
massive quiescent galaxies at $z\sim 2$, compact quiescent galaxies
likely
grow in size and mass at approximately the same rate as the galaxies
with dense cores. 
This result is consistent with
many previous theoretical and observational studies
(e.g., {Loeb} \& {Peebles} 2003; {Bezanson} {et~al.} 2009; {Naab} {et~al.} 2009; {Hopkins} {et~al.} 2010; {van Dokkum} {et~al.} 2010; {Oser} {et~al.} 2010;
Trujillo et al.\ 2011; {Newman} {et~al.} 2012; {Patel} {et~al.} 2013; {Szomoru} {et~al.} 2013; {Hilz} {et~al.} 2013).
It is also consistent with {Belli}, {Newman}, \& {Ellis} (2013), who made a similar
argument based on
the structural evolution of galaxies at constant velocity dispersion.
It is in conflict with studies that have suggested
that massive quiescent galaxies
evolve very little since $z\sim 2$  ({Carollo} {et~al.} 2013; {Poggianti} {et~al.} 2013),
and that
the apparent evolution is largely or entirely
due to the continuous addition of recently quenched galaxies to the
sample.\footnote{``Progenitor bias''; see {van Dokkum} \& {Franx} (1996, 2001)
for a description of this bias
in the context of 
samples selected by morphology.}

\section{Summary and Conclusions}
\label{conclusions.sec}

In this paper we have identified dense cores in galaxies out
to $z=2.5$, using data from the 3D-HST project augmented
by low redshift information from UltraVISTA and the Sloan
Digital Sky Survey. We find that the evolution of cores
with mass $\log(M_{\rm 1\,kpc}) \sim 10.5$
is well described by mild mass loss, suggesting
that their stars form a passive stellar population since
$z\sim 2.5$.  We note that mergers may also contribute to the evolution,
and that the effects of mass loss are sensitive to the assumption
that 100\,\% of the stellar ejecta mix with the hot halo gas.
At $z\sim 2.5$ the cores make up $\sim 50$\,\% of
the total mass of the galaxies that they are part of. At lower redshift
they make up a decreasing fraction of the total mass, and by $z=0$
they are embedded in large envelopes of stars with effective radii
$\sim 5$\,kpc. 

We focused on cores of a fixed high mass, but we note
that the evolution of the core mass function is
mass-dependent (see Fig.\ \ref{densfunc.fig}),
with low mass
cores showing strong positive evolution in their number density.
This mass dependence has also
been seen in the total mass function ({Marchesini} {et~al.} 2009) and
in the velocity dispersion function ({Bezanson} {et~al.} 2011). At low
masses star formation may lead to a relatively uniform build-up
of galaxies, with the stellar density increasing at all radii,
whereas at high masses galaxies are built up inside-out
(see {van Dokkum} {et~al.} 2013).

The negative mass evolution of the cores has consequences for the
interpretation of massive star forming galaxies at $z=1-2.5$ and
the evolution of quiescent galaxies, as discussed in \S\,\ref{buildup.sec}.
However,
we emphasize that not all massive galaxies have dense cores:
selecting on total mass produces
different samples than selecting on core mass, as is
obvious in Figs.\ \ref{massmass.fig} and \ref{massfrac.fig}.
Our conclusions only hold for galaxies with a dense core, and leave
open the possibility that massive galaxies with low core masses
have different evolutionary trajectories. It so happens that
by $z\sim 2$ our
selection mostly overlaps with the population of
massive, quiescent galaxies at that redshift, which is why
we can rule out several proposed models for their
evolution (see \S\,\ref{buildup.sec}).

We also find that, at fixed total mass and redshift,
the presence of a dense core
is a good predictor of quiescence and (perhaps
more interestingly) its absence is a nearly perfect predictor of
star formation (see Fig.\ \ref{massfrac.fig}).
The latter result is strikingly unambiguous: of 91
galaxies with $M_{\rm tot}>10^{11}$\,\msun\ and $M_{\rm
1\,kpc}<10^{9.5}$\,\msun\ only one is quiescent. Apparently the
presence of a dense
core is a ``non-negotiable'' requirement for
stopping star formation in massive galaxies.

Perhaps the most important result of this paper
is that the contribution of stars in dense cores
to the stellar mass
density of the Universe increases strongly with redshift, reaching
values of 10\,\% -- 20\,\% at $z\sim 2$ (\S\,\ref{contrib.sec}
and Fig.\ \ref{fractions.fig}b).
In light of this high fraction
we suggest that the formation of these cores is
an important aspect of star formation, galaxy formation,
and black hole formation at high redshift.

Interestingly it is not yet clear how this happened.
Near the end of their main star formation epoch,
prior to stellar mass loss, the cores were even more massive and compact than
at $z\sim 2$. The gas mass that was converted to stars inside
1\,kpc must have approached $10^{11}$\,\msun. Furthermore, this
gas must have arrived in the core without forming many stars at larger radii:
the quiescent descendants at $z\sim 2$ have  small
effective radii and no low surface brightness envelopes
(e.g., {Szomoru} {et~al.} 2010, 2013). Several mechanisms have been
proposed for creating very compact massive galaxies, such as
mergers ({Hopkins} {et~al.} 2008)  and disk instabilities ({Dekel} \& {Burkert} 2014).
However, reproducing the surface density profiles of the cores
has proven to be  challenging (see {Wuyts} {et~al.} 2010).
It will also be interesting to see whether models can be created that
simultaneously explain the existence of large, massive disks such as that
of M101 and of extremely compact cores of similar mass. Forming large
disks requires feedback and
significant angular momentum (e.g., {Guedes} {et~al.} 2011),
whereas forming dense cores requires rapid cooling
and a mechanism to lose angular momentum efficiently
({Sales} {et~al.} 2012; {Dekel} \& {Burkert} 2014).
Whatever the mechanism is for getting gas into the center, the core
mass will build up quickly when star formation begins. The adiabatic
enhancement discussed in Appendix \ref{lossaper.sec} should also
apply ``in reverse'': when mass is added to the center, the mass
within 1\,kpc will increase as $\sim (M'/M)^2$ due to adiabatic
contraction.

Whether forming dense cores have
been observed is a matter of debate.
As discussed in \S\,\ref{buildup.sec} star forming
galaxies with dense cores, such as those identified
by {Patel} {et~al.} (2013) and {Barro} {et~al.} (2013),
may not be forming the core itself but stars away from the center.
Spatially-resolved star formation maps 
(e.g., {F{\"o}rster Schreiber} {et~al.} 2011; {Nelson} {et~al.} 2013; {Wuyts} {et~al.} 2013), or spectroscopy to
determine the kinematics of the gas,
may provide more information on the location of star formation
in these objects.
Given the high metallicity of the centers of present-day elliptical
galaxies and the high densities, the
star forming cores must have had very large
amounts of absorption. They may be largely invisible in the optical and
near-IR, and possibly even at larger wavelengths.\footnote{Somewhat
akin to dragonflies, which are aquatic during their nymph stage.}
Studies of 
red, far-IR selected galaxies have shed some light on this issue
(e.g., {Tacconi} {et~al.} 2006, 2008; {Wang}, {Barger}, \& {Cowie} 2012; {Gilli} {et~al.} 2014).
The ``prototype'' would
be a dusty star forming galaxy with
a compact morphology and a gas dispersion that matches
the dispersion of present-day elliptical galaxies; such an object
has recently been identified (Nelson et al.\ 2014).

The main uncertainty in the analysis is the conversion of light
to mass. As discussed in Appendix A and elsewhere,
the systematic uncertainties are $\sim 0.1$\,dex, or half of the
observed evolution in the core mass. Stellar kinematics are
a crucial check on the mass measurements (see,
e.g., {Bezanson} {et~al.} 2013; {van de Sande} {et~al.} 2013; {Belli} {et~al.} 2013), although 
models for the structure of the galaxies and
their dark matter are required to interpret them.
Furthermore, we have
ignored radial gradients in $M/L$ ratio.
Our analysis shows that the core masses do not grow but the total masses
do, which means the stellar populations in the core are likely different
from those at larger radii. 
The available evidence suggests
that these gradients are generally small ({Szomoru} {et~al.} 2013), but it
is difficult to measure them at the relevant spatial scales: 1 kpc
corresponds to a single native WFC3 pixel. Spatially-resolved
studies of strongly-lensed galaxies with dense cores could address
this issue.

\begin{acknowledgements}
We thank the anonymous referee for constructive comments that improved
the manuscript. Support from STScI grant GO-12177 is gratefully acknowledged. 
The UltraVISTA catalog
is based on data products from observations made with ESO Telescopes at
the La Silla Paranal Observatory under ESO programme ID 179.A-2005 and
on data products produced by TERAPIX and the Cambridge Astronomy
Survey Unit on behalf of the UltraVISTA consortium.
\end{acknowledgements}

\begin{appendix}
\section{A.\ Tying the Wide-Field COSMOS Data to 3D-HST}
\label{uvista.sec}

In the main text
we augment the 3D-HST survey with
data from the UltraVISTA ({Muzzin} {et~al.} 2013b)
and ``Z\"urich'' ({Sargent} {et~al.} 2007) programs in the 1.5\,degree$^2$
COSMOS field. Here we compare total masses and core masses of objects
that are in common between the two surveys, and derive an offset
to place the wide-field data on the same system as the 3D-HST data.
We also use a third survey, the NEWFIRM Medium Band Survey
(NMBS; {Whitaker} {et~al.} 2011). We note that
none of these surveys are completely
independent; in particular, 3D-HST uses imaging
data from both the NMBS and from UltraVISTA.

\begin{figure*}[hbtp]
\epsfxsize=17cm
\epsffile[30 366 565 690]{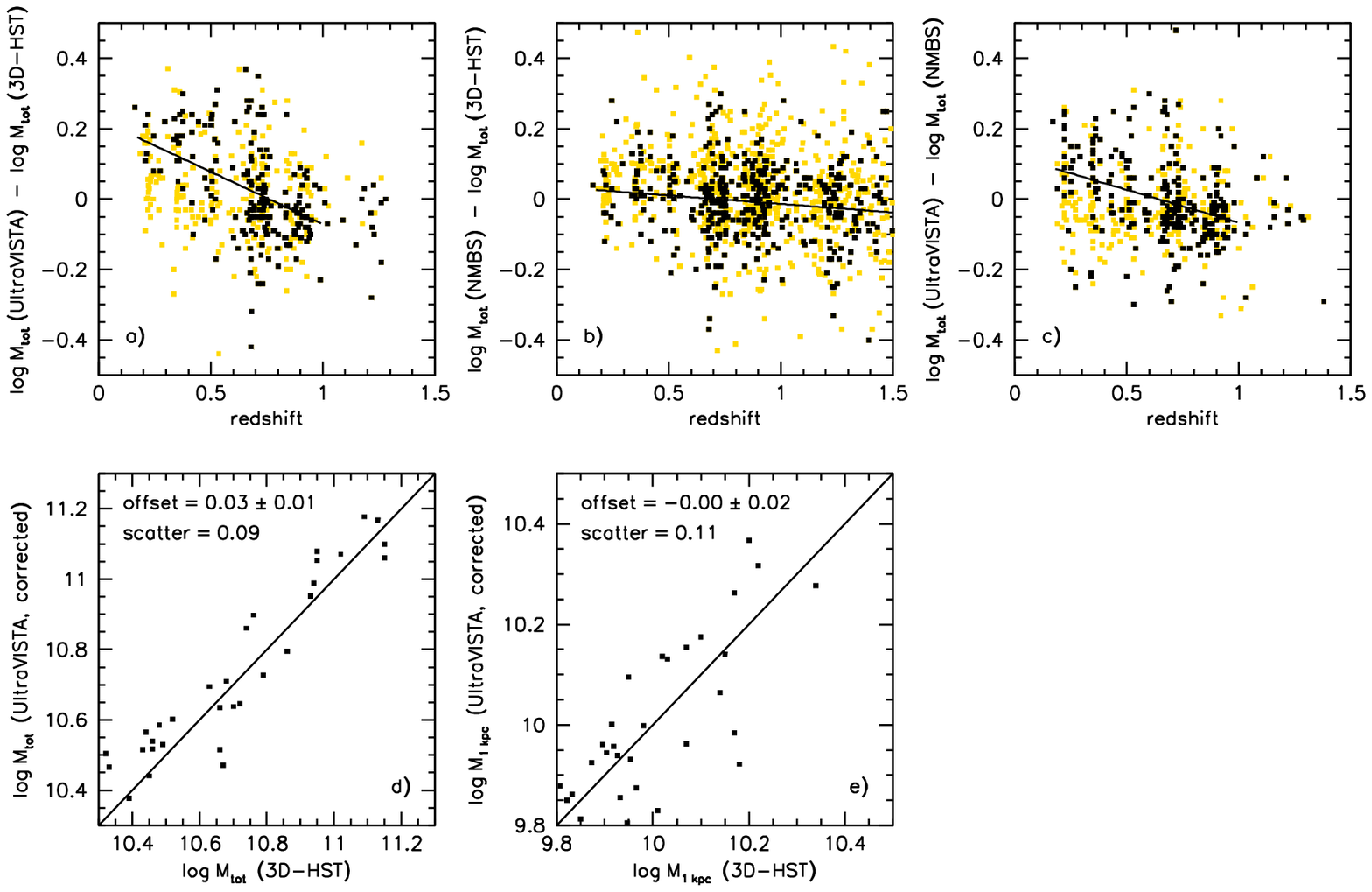}
\caption{\small
{\em a)} Comparison of total masses measured in UltraVISTA to those measured
in 3D-HST, for the same objects. Yellow dots are galaxies with
$9<\log(M_{\rm 1\,kpc})<9.8$; black dots are galaxies
with $\log(M_{\rm 1\,kpc})\geq 9.8$. The line is a fit to the black points.
{\em b)} Comparison between NMBS and 3D-HST. There is no evidence for
a systematic difference. {\em c)} Comparison between UltraVISTA and
NMBS. The same difference is evident as between UltraVISTA and 3D-HST:
the most massive galaxies at low redshift
have slightly higher masses in UltraVISTA. {\em d)} Mass-mass diagram
showing total masses in UltraVISTA, offset using the black line in
panel a, versus 3D-HST masses. {\em e)} Comparison of masses
within 1\,kpc, after offsetting the UltraVISTA masses to the 3D-HST
system. The two datasets agree within the uncertainties.
\label{compare.fig}}
\end{figure*}

The comparison sample is limited to objects in the 3D-HST/CANDELS
COSMOS field that have stellar mass measurements
from UltraVISTA and GIM2D structural parameters from the Z\"urich
catalog. 
Figure \ref{compare.fig}a shows the difference in stellar mass
between 3D-HST and UltraVISTA, as a function of redshift. Yellow points
are objects with $9<\log(M_{\rm 1\,kpc})<9.8$, and black points are
objects with $\log(M_{\rm 1\,kpc})\geq 9.8$. The two surveys produce
consistent masses for the vast majority of objects: the median
difference for the yellow points is only $0.01$. However, the
subset of galaxies with 
high core masses (and the highest total masses) and low
redshifts show an offset. The black line is a 
fit of the form
\begin{equation}
\label{uvista.eq}
\log(M_{\rm tot}) ({\rm UltraVISTA}) - \log(M_{\rm tot}) ({\rm 3D-HST}) =
 0.23 - 0.30 z.
\end{equation}
This fit is valid for $\log M_{\rm 1\,kpc}>9.8$ and $0.2<z<1$.

In Fig.\ \ref{compare.fig}b we show the difference between masses in 3D-HST
and in the NMBS survey ({Whitaker} {et~al.} 2011). The NMBS is a ground-based
$K$-selected survey, like UltraVISTA, and it uses similar photometric
bands.\footnote{The main difference is that NMBS uses medium-bandwidth near-IR
filters, which leads to improved photometric redshifts at $z\gtrsim 1$. NMBS
covers a $6\times$
smaller part of the COSMOS field than UltraVISTA.}
There is no systematic offset between 3D-HST stellar masses and NMBS
stellar masses. This conclusions also applies to the most massive
galaxies (black points): the black line is a fit to the most massive
objects, and it is within $0.03$\,dex of zero at all redshifts.
Panel c compares masses in UltraVISTA to those in NMBS.
As expected from panels a and b, we find that UltraVISTA and NMBS
have systematically different masses for massive galaxies at low
redshift. This subtle redshift-dependent effect is not evident
in Fig.\ 8 of {Muzzin} {et~al.} (2013b); it is present when that
figure is remade for the most massive
galaxies in the redshift range $0<z<0.5$, at a level
consistent with Fig.\ \ref{compare.fig}c.

Based on these comparisons we apply an offset to
the UltraVISTA masses, using
the relation in Eq.\ \ref{uvista.eq}.
We stress that we do not know whether
the masses in UltraVISTA or in 3D-HST are closer to the
correct values; we simply adopt
the 3D-HST system as our default and add the applied offset in quadrature
to our error budget.
In Fig.\ \ref{compare.fig}d
we show the relation between the total masses in UltraVISTA and in
3D-HST after applying the offset,
for galaxies with $\log(M_{\rm 1\,kpc})>9.8$ and 
$0<z<0.5$. The difference is now close to zero, as expected.
The scatter is 0.09\,dex. Assuming that the  errors that cause
this scatter are of similar magnitude
and independent in both surveys (which is an oversimplification),
we infer that the error in an individual mass measurement is
$0.09/\sqrt{2} = 0.06$\,dex.

In Fig.\ \ref{compare.fig}e
we compare the derived core masses for these same galaxies. The
core masses were calculated from the total masses and the structural
parameters of the galaxies (see \S\,\ref{select.sec}). These structural
measurements are completely independent: for 3D-HST they
are measured using GALFIT ({Peng} {et~al.} 2002) from WFC3 $J_{125}$ images
(in this redshift range), whereas for COSMOS they were measured using
GIM2D ({Simard} {et~al.} 2002) from ACS $I_{814}$ images. The core masses
have an offset of $0.00 \pm 0.02$ dex, which means the two surveys
produce consistent core masses after the offset that
we applied to the total masses in UltraVISTA. The random error
for a single measurement is approximately $0.08$\,dex.
The core mass function
(i.e., the number
density of galaxies as a function of their mass within 1\,kpc)
in UltraVISTA is shown
in Fig.\ \ref{densfunc.fig}b in the main text; if we had not
applied the mass offset
the core mass function in this panel would be shifted by $+0.09$ dex.

\begin{figure*}[hbtp]
\epsfxsize=12.5cm
\epsffile[-250 168 572 696]{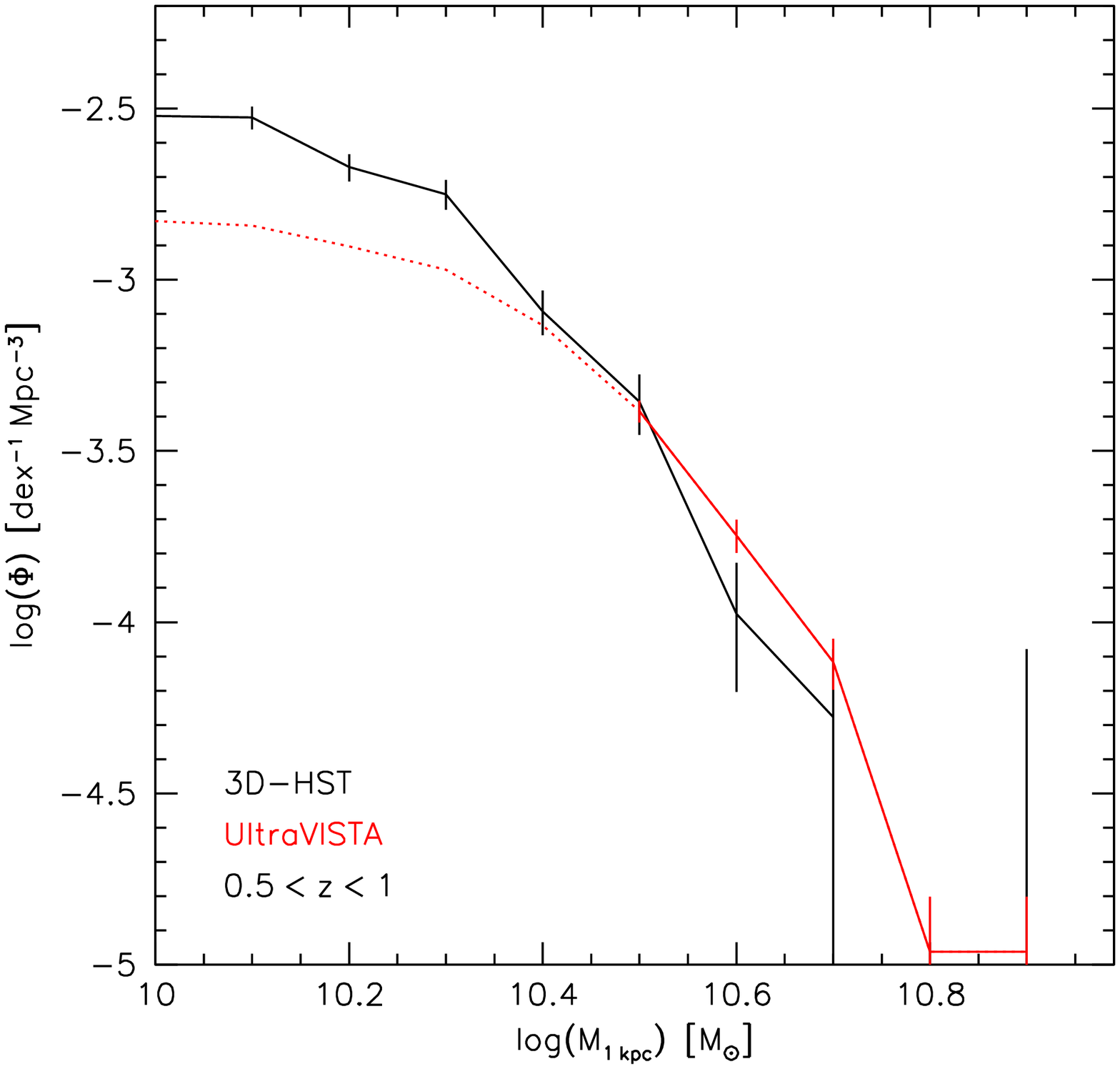}
\caption{\small
Number density of galaxies as a function of their core mass, for $0.5<z<1$
and data from 3D-HST (black line) and from the
UltraVISTA/Z\"urich wide-field
survey of the COSMOS field (solid red line). The
broken red line indicates the regime where the Z\"urich magnitude
limit leads to incompleteness. The two surveys are in good agreement
in the regime where they are both complete.
\label{compare2.fig}}
\end{figure*}

As a further check we compare the core mass function
from the wide field COSMOS data to that derived from 3D-HST.
We cannot compare the redshift range $0<z<0.5$
as the 3D-HST data do not sample enough volume (which is why we turned
to UltraVISTA/Z\"urich for this redshift range). However, we can do this
comparison at $0.5<z<1$. Here we are constrained by the $I=22.5$ limit
of the Z\"urich morphological catalog; this  
limit implies that the completeness drops below 80\,\% for core masses
$\log (M_{\rm 1\,kpc})\lesssim 10.5$. We show the comparison
between 3D-HST and UltraVISTA/Z\"urich in Fig.\ \ref{compare2.fig}.
The data are in excellent agreement for $\log (M_{\rm 1\,kpc})>10.4$.

\section{B.\ Effect
of Random Errors on the Evolution of the Core Mass Function}
\label{effect_errors.sec}

Due to the steepness of the core mass function random errors
can lead to an artifical increase in the number density
of the cores with the highest masses (see, e.g., {Bezanson} {et~al.} 2011, for
an analysis of this effect on the velocity dispersion function).
If the random errors are a function of redshift, such that the highest
redshift data suffer from the largest errors, they might explain part
or all of the observed evolution in the core mass function.

We analyze the effects of random errors in the following way. The
observed core masses of individual galaxies
in the Sloan Digital Sky Survey (SDSS) are perturbed using a
log-normal probability distribution of width $s$. Then, the
core mass function is constructed using these perturbed masses and
compared to the observed core mass function at $2<z<2.5$. The value
of $s$ is related to
the errors in the core masses at high redshift, $e_h$,
and the errors at low redshift, $e_l$, through
$e_h = (s^2 + e_l^2)^{0.5}$.

\begin{figure*}[hbtp]
\epsfxsize=8.5cm
\begin{center}
\epsffile[29 179 568 690]{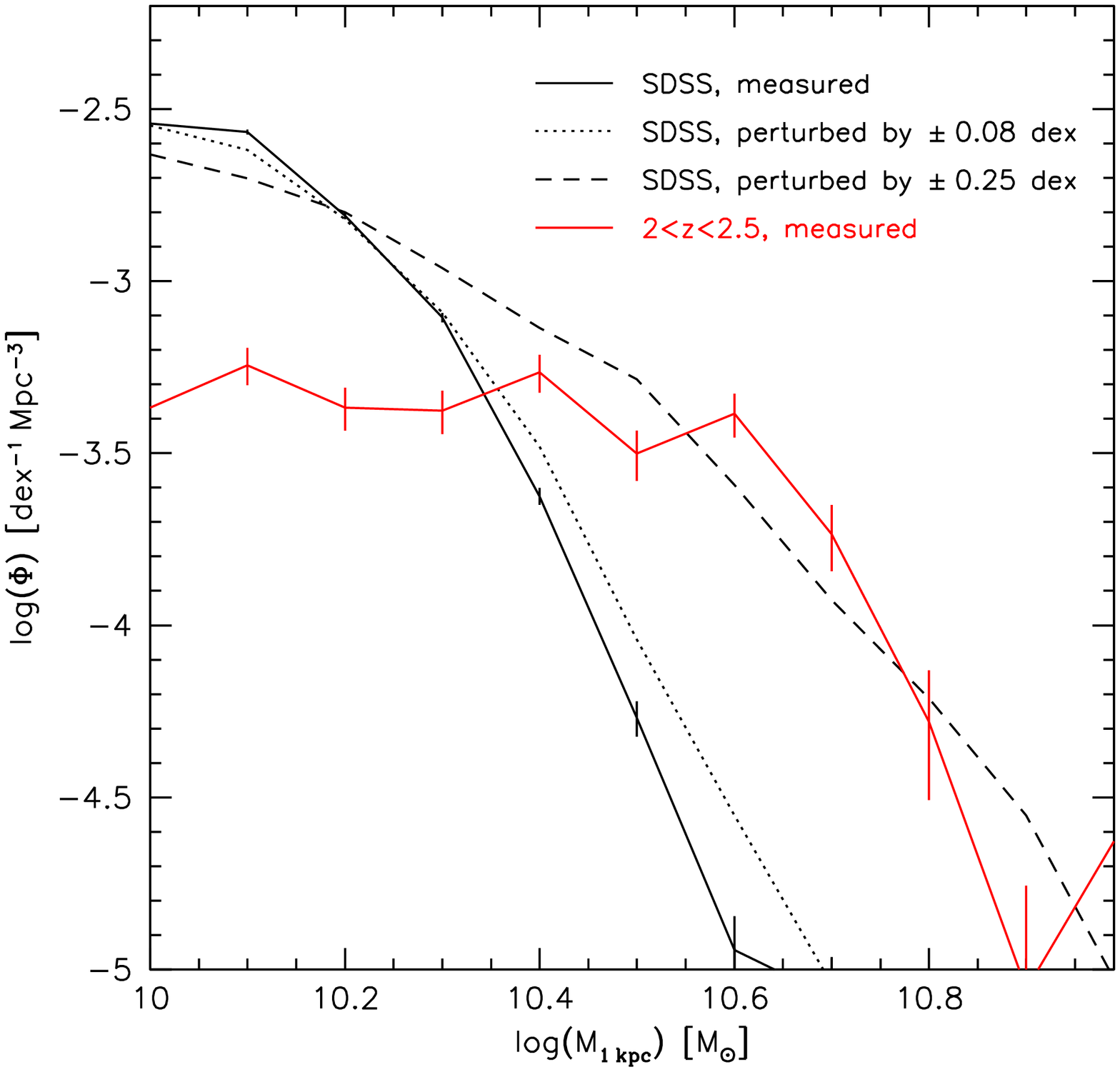}
\end{center}
\caption{\small
Effect of random errors on the evolution of the core mass function.
The dotted line shows the SDSS core mass function after perturbing the
core masses by a Gaussian of
width $s=0.08$\,dex, which is the empirically-determined uncertainty
in individual measurements in the 3D-HST survey. 
Random errors in this range cannot explain the observed difference between the
core mass function at $2<z<2.5$ and at low redshift.
Errors of $\pm 0.25$\,dex (dashed line) would be required to bring the
two functions into agreement.
\label{corefunc_error.fig}}
\end{figure*}

In Fig.\ \ref{corefunc_error.fig} we show the effects on the observed
core mass function for two values of $s$, 0.08\,dex and 0.25\,dex.
The value of 0.08 is derived from the analysis in Appendix \ref{uvista.sec},
where we show that this is the approximate random error $e_h$
in an individual measurement in
3D-HST. Assuming that the SDSS measurements have
no error and hence $e_l\sim 0$, we find $s\sim 0.08$\,dex.
This can be regarded as a ``maximum plausible'' error, as there is no
a priori reason why the SDSS measurements should have a much
smaller error
than the high redshift data. It is clear
from the dotted line in Fig.\ \ref{corefunc_error.fig} that random
errors of this magnitude
have little effect on the inferred evolution of the
core mass function. To bring the SDSS core
mass function into agreement with the
observed core mass function at high redshift, the 
random errors at high redshift would have to be $\sim 0.25$\,dex
greater than those in the SDSS.

\section{C.\ Field-to-Field Variations}
\label{cosvar.sec}

Galaxies with dense cores are rare and presumably live in massive dark
matter halos; it is therefore a concern that the results in this paper are
driven by one or two
overdense or underdense regions of the Universe. This is
a particular concern for the analysis of the evolution of the number
density of massive cores (\S\,\ref{ndens.sec}).
Fortunately, the 3D-HST/CANDELS survey covers five survey
fields in completely different regions of the sky, and we can test
whether the densities in the five fields are similar.

\begin{figure*}[hbtp]
\epsfxsize=8.5cm
\begin{center}
\epsffile[23 175 575 700]{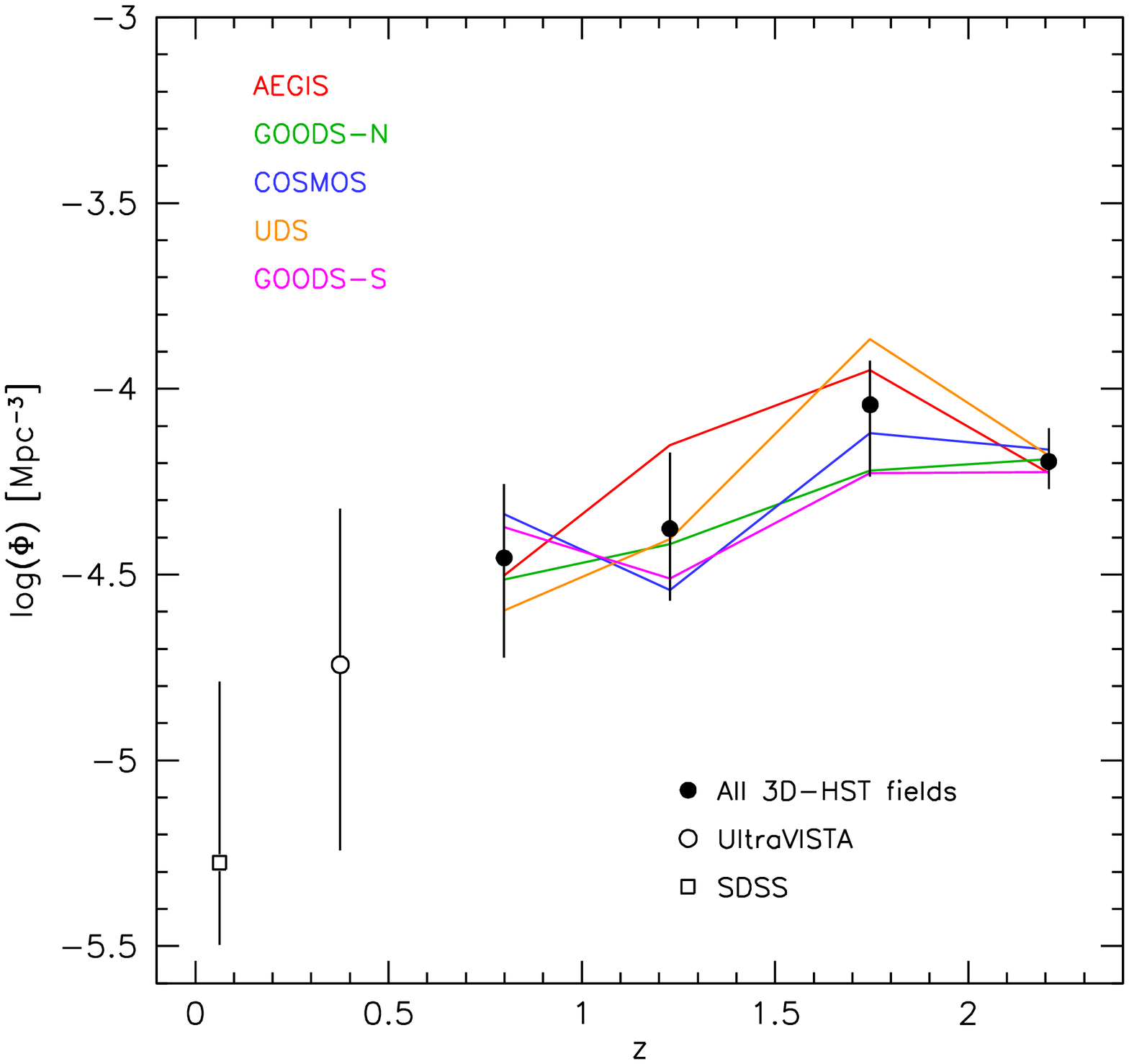}
\end{center}
\caption{\small
Field-to-field variation in the cumulative number density
of dense cores. Black points are identical to those plotted
in Fig.\ \ref{dens_evo.fig} in the main text. Colored lines show
the evolution as measured in each of the five 3D-HST/CANDELS
fields. The evolution is consistent, and the error due to field-to-field
variation (``cosmic variance'') is small compared to other sources
of error. 
\label{cosvar.fig}}
\end{figure*}

Figure \ref{cosvar.fig} is a repeat of Fig.\ \ref{dens_evo.fig}, and
shows the evolution of the cumulative number density of galaxies
with $\log(M_{\rm 1\,kpc})>10.5$. Colored lines show the evolution
as measured from the five individual 3D-HST/CANDELS fields. The
individual fields show the same evolution as the five fields combined,
and there is no single field that significantly alters the average
at a particular redshift. The scatter between the fields is $\approx 0.15$\,dex
for the three lowest redshift bins and only $0.04$\,dex at $z=2-2.5$.
We conclude that the error in the mean due to cosmic variance
is approximately
$0.15/\sqrt{5} \approx 0.07$\,dex, much smaller than the
errors due to mass uncertainties (see main text).
We also infer that
there are no large differences in the absolute mass calibrations between
the five fields, as they would ``translate'' into large variation in
the normalization of the five curves.

\section{D.\ Pairs of Galaxies with Dense Cores}
\label{pair.sec}

\begin{figure*}[hbtp]
\epsfxsize=17cm
\epsffile[30 206 692 406]{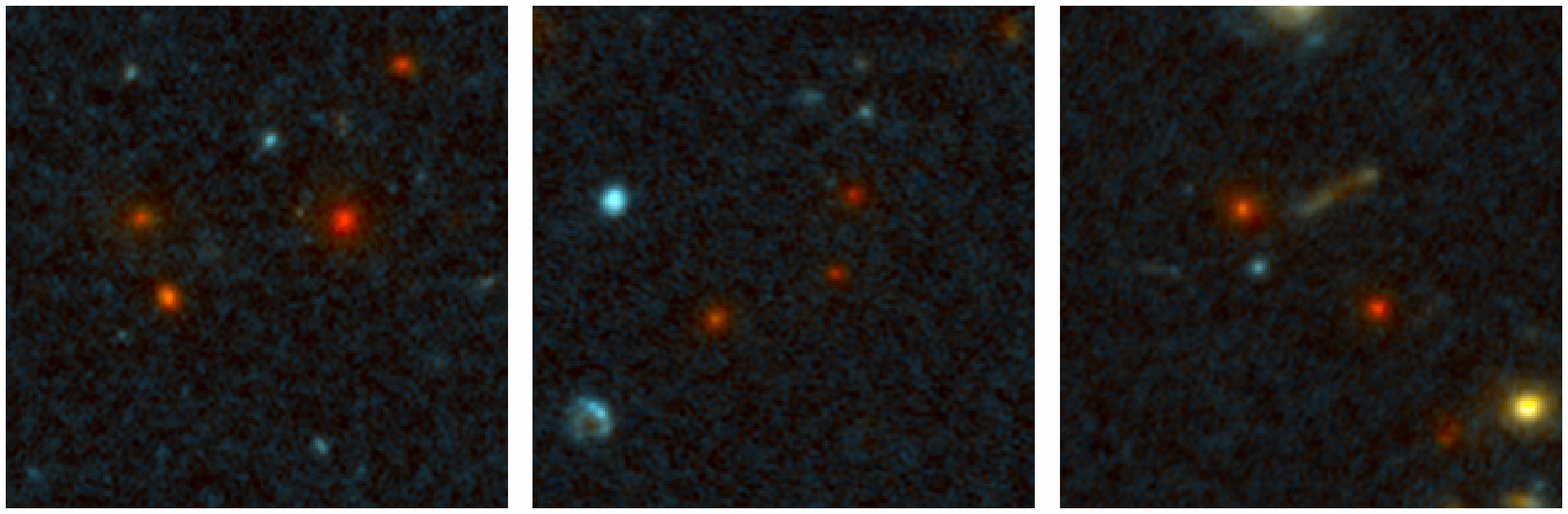}
\caption{\small
The only
pairs of galaxies with dense cores in the entire 3D-HST survey:
a pair at $z\approx 1.71$ in the AEGIS field (left), a pair
at $z\approx 2.31$ in COSMOS (middle), and a pair
at $z\approx 2.07$ in UDS  (right). Interestingly, all three pairs have
other similarly red galaxies in their vicinity.
\label{pairs.fig}}
\end{figure*}

As discussed in \S\,\ref{mergers.sec} there are only three  pairs
of galaxies with dense cores  in the
3D-HST/CANDELS fields, out of
a parent population of 267 galaxies with 
$\log (M_{\rm 1\,kpc})>10.5$. One pair is in the COSMOS field,
one in the AEGIS field, and one in the UDS field. No core-core
pairs with projected separations $d<43$\,kpc
are found in either of the GOODS fields. Color images of the three
pairs are shown in Fig.\ \ref{pairs.fig}.

\section{E.\ The Effect of Mass Loss on the Mass Within 1\,kpc}
\label{lossaper.sec}

Stellar mass loss can affect the mass within a fixed aperture of 1\,kpc in
two ways: directly through the mass that is lost in the winds, and
indirectly through the adiabatic expansion that follows the change in
mass. As discussed in the main text, the latter effect is only
important if 100\,\% of the stellar ejecta mix with the hot halo
gas. Here we calculate the total effect on $M_{\rm 1\,kpc}$,
the stellar mass within 1\,kpc, under the assumption that all the
material is heated and is diffusely distributed in the hot halo.
The galaxy's effective radius
will then increase as
\begin{equation}
\label{expand.eq}
r_e'/r_e \sim (M'/M)_{\rm tot}^{-1}
\end{equation}
due to adiabatic expansion ({Hills} 1980; {Fan} {et~al.} 2008; {Ragone-Figueroa} \& {Granato} 2011). The
effect on the mass within 1\,kpc therefore
depends on the structure
of the galaxy.  If most of the total mass
is within 1\,kpc to begin with the effect is negligible and
$(M'/M)_{\rm 1\,kpc} \sim (M'/M)_{\rm tot}$.
On the other hand, if the galaxy has a density profile that is nearly
constant with radius (and therefore $r_e \gg 1$\,kpc), the
mass inside 1 kpc will decrease as $(M'/M)_{\rm 1\,kpc} \sim
(M'/M)_{\rm tot} \times (r_e'/r_e)^{-3} \sim (M'/M)_{\rm tot}^4$.

\begin{figure*}[htbp]
\epsfxsize=8.6cm
\begin{center}
\epsffile[23 175 575 700]{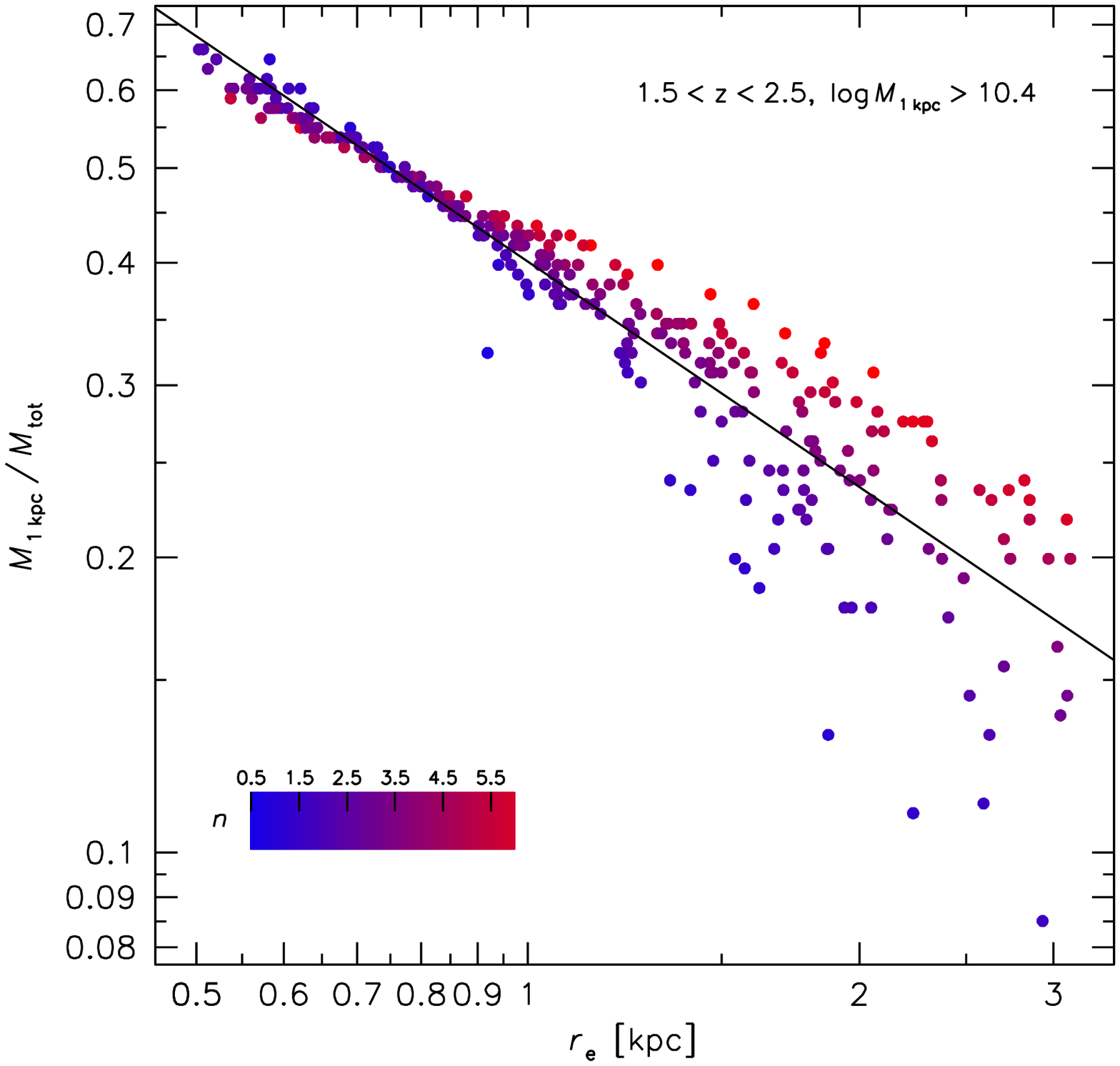}
\end{center}
\caption{\small
Relation between the effective radius and the fraction of the total
mass that is in a dense core, for galaxies with $1.5<z<2.5$ and $\log
(M_{\rm 1\,kpc})>10.5$. Objects are color-coded by their Sersic index,
going from blue (low $n$) to red (high $n$). The line is a fit to
all the points, and has a slope of $-0.8$. This relation is used to
approximate the effect of adiabatic expansion on the mass enclosed
within 1\,kpc.
\label{adia.fig}}
\end{figure*}

In practice the effect on $M_{\rm 1\,kpc}$ will be in between
these two extremes. We determined the change in $M_{\rm 1\,kpc}$
empirically using the actual galaxies in the 3D-HST survey.
Figure\ \ref{adia.fig} shows the ratio between $M_{\rm 1\,kpc}$
and $M_{\rm tot}$ as a function of the effective radius, for
galaxies with $\log(M_{\rm 1\,kpc})>10.4$, $1.5<z<2.5$,
and $-0.3<\log(r_e)<0.5$. The slope of
the relation ranges between $-0.6$ and $-1.3$
depending on the {Sersic} (1968)
index. The average relation, shown by the solid line,
is $M_{\rm 1\,kpc}/M_{\rm tot} \propto r_e^{-0.8}$. 

We infer that the total effect of stellar
mass loss on the mass within 1 kpc is
\begin{eqnarray}
\label{adia_app.eq}
(M'/M)_{\rm 1\,kpc} & \sim & (M'/M)_{\rm tot} \times (r_e'/r_e)^{-0.8} \nonumber \\
  & \sim & (M'/M)_{\rm tot}^{1.8}.
\end{eqnarray}
We note that this relation assumes that there are no other
sources of mass loss than stellar evolution.
If there are other sources of mass loss, such
as AGN-driven winds (e.g., {Fan} {et~al.} 2008), the
adiabatic component of Eq.\ \ref{adia_app.eq} may be larger.

\section{F.\ Core Mass Function Using Major Axis Effective Radii}
\label{major.sec}

In the default analysis in the paper we use the circularized effective
radius to deproject the best-fitting {Sersic} (1968) profiles and
measure the core mass. As noted in \S\,\ref{other.sec}, the
circularization may introduce biases if the mean axis ratio of galaxies
evolves with redshift: at fixed circularized effective radius,
highly flattened galaxies have less
mass in a sphere of 1 kpc than spherical galaxies.
We determined the importance of this effect by repeating the
deprojection, now using the major axis effective radius rather than
the circularized effective radius. The major axis effective radius
is always larger than the circularized one, and so the inferred
core masses decrease; the question is whether this decrease is
dependent on redshift.

\begin{figure*}[hbtp]
\epsfxsize=16.5cm
\epsffile[15 326 532 660]{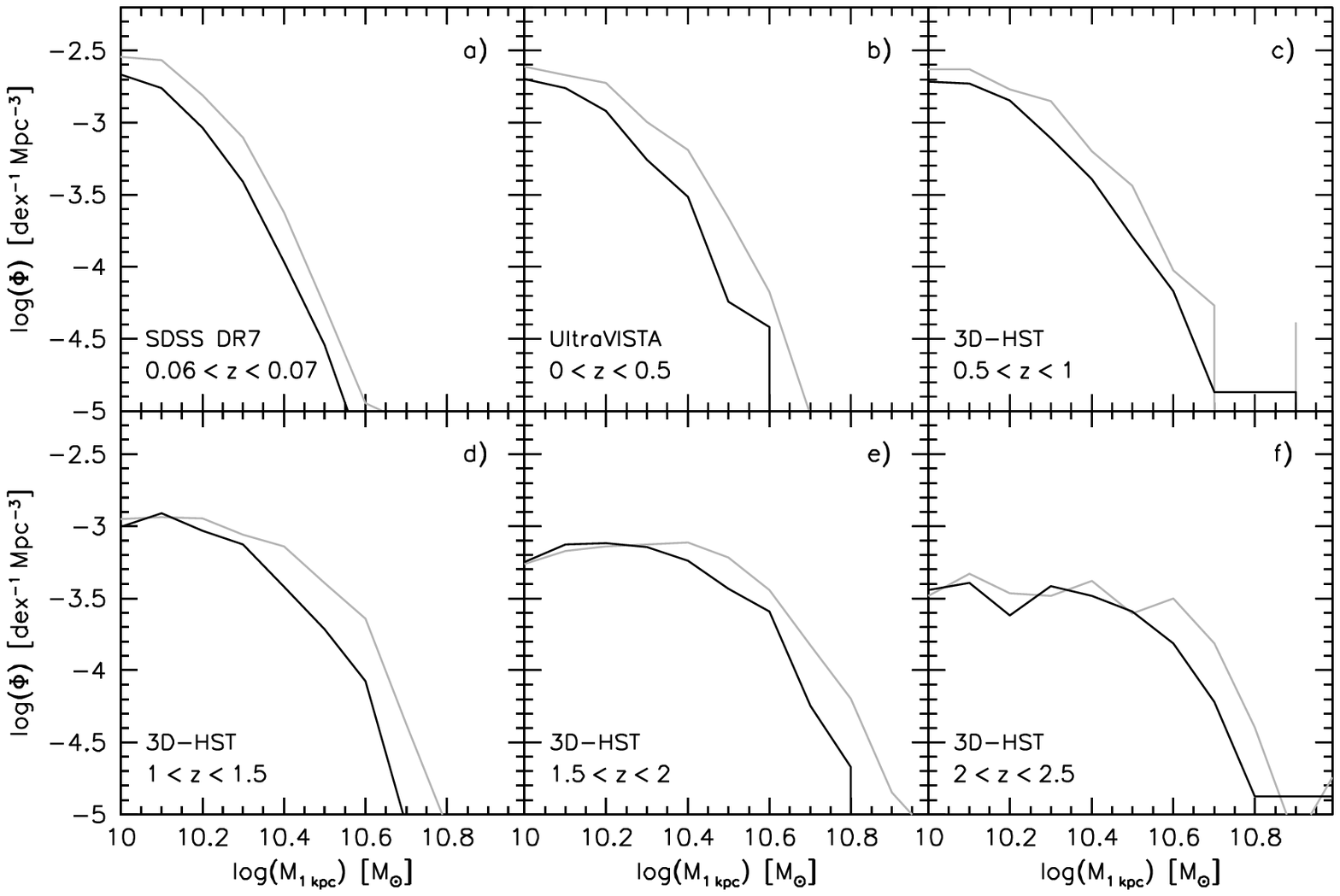}
\caption{\small
Evolution of the core mass function, using the major axis effective
radius in the deprojection rather than the circularized effective radius
(black lines). The default deprojections used in the paper are shown
in grey. There is a constant offset in mass, as expected, but no significant
redshift-dependent difference between the grey and black curves.
\label{densfunc_a.fig}}
\end{figure*}

The core mass function, as derived using the major axis effective radius,
is shown by the black lines in Fig.\ \ref{densfunc_a.fig}. It is compared
to the original, circularized measurements in grey. There is no
significant redshift-dependent effect, and we conclude that systematic
changes in axis ratio are not the cause of the apparent negative evolution
in the number density of dense cores.

\end{appendix}


\end{document}